\documentclass[
superscriptaddress,
amsmath,amssymb,
aps, 
prd,
twocolumn,nofootinbib
]{revtex4-1}

\usepackage[dvipsnames]{xcolor}
\usepackage{graphicx}% Include figure files
\usepackage{dcolumn}% Align table columns on decimal point
\usepackage{cases}
\usepackage{comment}
\usepackage{ulem}
\usepackage{bm}% bold math
\usepackage[colorlinks,urlcolor=Maroon,linkcolor=Maroon,anchorcolor=blue,citecolor=MidnightBlue]{hyperref} %Use hyperlinks and set their colors
\allowdisplaybreaks[4]

\begin{document}

\title{%Relativistic perturbation theory and environmental effects on superradiant instabilities
Revisiting environmental effects on black hole quasibound-state spectra with relativistic perturbation theory
}% 

\author{Yin-Da Guo}
\email{yinda.guo@mail.sdu.edu.cn}
\affiliation{Key Laboratory of Particle Physics and Particle Irradiation (Ministry of Education),\\Institute of Frontier and Interdisciplinary Science, \\Shandong University, Qingdao 266237, China}
\affiliation{CENTRA, Departamento de Física, Instituto Superior Técnico – IST, Universidade de Lisboa – UL, Avenida Rovisco Pais 1, 1049–001 Lisboa, Portugal}

\author{Qi-Xuan Xu}
\email{qixuan.xu@tecnico.ulisboa.pt}
\affiliation{CENTRA, Departamento de Física, Instituto Superior Técnico – IST, Universidade de Lisboa – UL, Avenida Rovisco Pais 1, 1049–001 Lisboa, Portugal}

\author{Richard Brito}
\email{richard.brito@tecnico.ulisboa.pt}
\affiliation{CENTRA, Departamento de Física, Instituto Superior Técnico – IST, Universidade de Lisboa – UL, Avenida Rovisco Pais 1, 1049–001 Lisboa, Portugal}

\author{Enrico Cannizzaro}
\email{enrico.cannizzaro@tecnico.ulisboa.pt}
\affiliation{CENTRA, Departamento de Física, Instituto Superior Técnico – IST, Universidade de Lisboa – UL, Avenida Rovisco Pais 1, 1049–001 Lisboa, Portugal}

\date{\today}% It is always \today, today,
             %  but any date may be explicitly specified

\begin{abstract}
We present a relativistic framework for computing corrections to the eigenfrequency spectrum of a massive scalar field in perturbed black-hole spacetimes, including first-order shifts to decay rates and second-order mode-mixing effects. We also clarify the regime of validity of non-relativistic treatments and show that the accuracy of completeness-based descriptions is limited, highlighting the non-Hermitian nature of the spectrum. Using galactic halos and accretion disks as physically motivated perturbations, we benchmark the relativistic perturbative predictions to the eigenfrequency shifts against non-perturbative numerical solutions. %As a byproduct, we also show that these environments induce only mild corrections to the spectrum for realistic astrophysical values. 
We also present first-order relativistic eigenfrequency shifts induced by binary companions, whose potentially stronger impact on superradiant dynamics of massive scalar fields around spinning black holes motivates future dedicated analyses. Our results suggest that previous estimates of the termination of superradiance due to binary companions and disks should be revisited within a relativistic framework.
\end{abstract}

\maketitle

%%%%%%%%%%%%%%%%%%%%%%%%
\section{Introduction}
%%%%%%%%%%%%%%%%%%%%%%%%
Similarly to how protons and electrons bind through electromagnetic interactions, black holes (BHs) and ultralight bosons can bind gravitationally to form long-lived gravitational atoms~\cite{Ternov:1978gq,Detweiler:1980uk,Dolan:2007mj}. Since bosons are not subject to Pauli blocking, a large number of particles can occupy the same energy level. In particular, occupation numbers can grow exponentially through superradiance, which extracts rotational energy from the BH leading to the formation of macroscopic boson clouds. This mechanism operates when the frequency $\omega$ of the boson field satisfies $\omega<m\Omega_\mathrm{H}$, where $m$ is the azimuthal number of the field and $\Omega_\mathrm{H}$ is the angular velocity of the event horizon. For a comprehensive review of superradiance, see Ref.~\cite{Brito:2015oca}.

Gravitational atoms stand out as promising systems to detect new fundamental bosons, and have been widely studied in the literature for scalar~\cite{Ternov:1978gq,Zouros:1979iw,Detweiler:1980uk,Cardoso:2005vk,Konoplya:2006br,Dolan:2007mj,Arvanitaki:2009fg,Arvanitaki:2010sy,Dolan:2012yt,Yoshino:2013ofa,Brito:2014wla,Arvanitaki:2014wva,Arvanitaki:2016qwi,Brito:2017wnc,Brito:2017zvb,Baumann:2018vus,Cardoso:2018tly,Baumann:2019eav,Baumann:2021fkf,Baumann:2022pkl,Tong:2022bbl,Brito:2023pyl,Takahashi:2023flk,May:2024npn,Zhu:2024bqs},
vector~\cite{Rosa:2011my,Pani:2012bp,Dolan:2018dqv,East:2017mrj,East:2017ovw,Baryakhtar:2017ngi,East:2018glu,Cardoso:2018tly,Frolov:2018ezx,Baumann:2019eav,Jia:2023see}
and tensor~\cite{Brito:2013wya,Brito:2020lup,Dias:2023ynv,East:2023nsk} fields. Most of these studies focus on bosonic clouds evolving around isolated Kerr BHs, where the superradiant spectrum and instability rates are by now well understood. However, realistic astrophysical BHs are not ``clean'' systems, and it is natural to ask whether these results remain robust in the presence of external perturbations, such as surrounding matter or a binary companion. Indeed, since superradiant growth rates are typically very small, even weak external perturbations may affect the instability. In this context, Ref.~\cite{Tong:2022bbl} argued that the gravitational potential of a companion can completely terminate superradiance by turning unstable modes into stable ones. The proposed mechanism relies on the breaking of axisymmetry, which induces mode mixing between superradiant states ($m>0$) and decaying ones ($m\leq0$). A similar mechanism was recently investigated in Ref.~\cite{Li:2026gup}, where the authors considered perturbations induced by a tilted accretion disk misaligned with the BH spin.

However, these studies rely on a non-relativistic treatment of the system. In the relativistic problem, the BH horizon renders the system intrinsically dissipative, leading to quasibound states with complex frequencies whose imaginary parts encode decay or superradiant growth through horizon fluxes. In the non-relativistic approximation, this dissipative boundary condition is replaced by regularity at the origin, reducing the problem to a Hermitian one with a purely real spectrum, exactly as in the hydrogen atom of quantum mechanics. As a consequence, dissipative effects are not captured and can only be introduced phenomenologically. This limitation may become particularly important when perturbations induce mode mixing, since the distinction between growth and damping originates precisely from the dissipative nature of the problem.

A fully relativistic framework for computing eigenfrequency corrections of quasibound states was recently developed in Ref.~\cite{Cannizzaro:2023jle}. This formalism is based on a bilinear form yielding a mode-orthogonality relation that consistently captures near-horizon physics, extending previous work on gravitational perturbations~\cite{Green:2022htq}. In this work we adopt and extend this framework to systematically study dissipative corrections to the spectrum. We show that already at first order in perturbation theory, the framework predicts an imaginary frequency shift, which is absent in the hydrogenic treatment. We then extend the formalism to second order, where mode mixing naturally arises. We show that hydrogenic treatments fail to completely capture the full relativistic dynamics, while relativistic approaches based on the completeness of quasibound states are accurate only within a limited regime, providing evidence that quasibound states do not form a complete basis.

We apply the framework to a set of physically motivated perturbations, including galactic halos, accretion disks, and binary companions, computing first-order relativistic corrections and benchmarking the predictions against numerical solutions whenever possible. Our results show that realistic halo and disk environments induce only mild corrections to the Schwarzschild and Kerr quasibound state frequencies. Using the recent results of Ref.~\cite{Lestingi:2026peq}, which show how to compute higher-order corrections, we then focus on the halo case to investigate second-order effects and completeness properties in a controlled setting, leaving the binary problem for future dedicated analyses.

This paper is organized as follows. In Sec.~\ref{sec:scalar_superradiance}, we provide a brief review of scalar superradiance. In Sec.~\ref{sec:environments}, we introduce the environmental configurations considered in this work, namely galactic halos, accretion disks, and binary companions. In Sec.~\ref{sec:perturbation_theory}, we present the perturbative frameworks used to compute frequency shifts. We derive the first-order shifts induced by astrophysical environments, using halos and accretion disks to benchmark the perturbative frameworks against numerical solutions, and then apply the formalism to binary companions.
In Sec.~\ref{sec:second_order}, we use the halo case to examine second-order perturbation theory and the validity of the completeness assumption for quasibound modes. Finally, Sec.~\ref{sec:summary} summarizes our main results and provides a brief discussion.

Throughout the paper, we adopt geometrized units $G=c=1$.

\section{Scalar field on a Kerr background}
\label{sec:scalar_superradiance}

The Kerr metric, characterized by the BH mass $M_\mathrm{BH}$ and angular momentum $J_\mathrm{BH}$, can be written in Boyer–Lindquist coordinates~\cite{Boyer:1966qh},
%%%%%=====
\begin{align}\label{eq:metric_kerr}
    \begin{split}
        ds^2 = & - \left(1-\frac{2M_\mathrm{BH}r}{\Sigma}\right) dt^2 - \frac{4 M_\mathrm{BH} a r \sin^2\theta}{\Sigma}dtd\varphi 
        \\
        & \hspace{-1cm} + \frac{\Sigma}{\Delta} dr^2 + \Sigma d\theta^2 + \frac{\sin^2\theta}{\Sigma}\left[(r^2+a^2)^2 - \Delta a^2 \sin^2\theta\right] d\varphi^2,
    \end{split}
\end{align}
%%%%%
where
%%%%%=====
\begin{align}
    a &= J_\mathrm{BH}/M_\mathrm{BH},\\
    \Sigma &= r^2 + a^2 \cos^2\theta, \\
    \Delta &= r^2 + a^2 - 2M_\mathrm{BH}r.
\end{align}
%%%%%
The inner (minus sign) and outer horizons (plus sign) are located at $r_\pm = M_\mathrm{BH} \pm \sqrt{M_\mathrm{BH}^2-a^2}$. The angular velocity of the outer horizon is given by $ \Omega_\mathrm{H} =a/(2M_\mathrm{BH}r_+)$.

We consider a test, free, complex scalar field $\Phi$ in the background of a Kerr or Schwarzschild BH. The corresponding leading-order Lagrangian density is
%%%%%=====
\begin{align}\label{eq:Lagrangian}
  \mathcal{L}^{(0)} = - g^{(0)\mu\nu} \nabla_{\mu}^{(0)}\Phi^{(0)*} \nabla^{(0)}_{\nu}\Phi^{(0)} - \mu_\mathrm{s}^2 \Phi^{(0)*}\Phi^{(0)},
\end{align}
%%%%% 
where $\hbar\mu_\mathrm{s}$ denotes the mass of the scalar field and $g^{(0)}_{\mu\nu}$ is the Kerr or Schwarzschild ($a=0$) metric. Here and in the following, the superscript ${(0)}$ denotes quantities computed in a Kerr/Schwarzschild background. 

The corresponding equation of motion is the Klein-Gordon equation:
%%%%%=====
\begin{align}\label{eq:KGeq}
  (\nabla^{(0)\mu} \nabla^{(0)}_{\mu} - \mu_\mathrm{s}^2) \Phi^{(0)} = 0.
\end{align}
%%%%% 
Given a physically motivated set of boundary conditions, namely ingoing waves at the future horizon and exponentially-decaying solutions at spatial infinity, the eigenfrequencies of this equation are, in general, complex. We denote its real and imaginary parts by $E_{nlm}^{(0)}$ and $\Gamma_{nlm}^{(0)}$, respectively, depending on the overtone number $n$, the angular number $l$, and the azimuthal number $m$. For comparison with the hydrogen atom spectrum, it is convenient to introduce $\bar{n} = n + l + 1$. For later use, we denote the eigenfrequency as $\omega_{nlm}^{(0)} \equiv E_{nlm}^{(0)} + i\Gamma_{nlm}^{(0)}$. The corresponding eigenfunction admits the separable form~\cite{Dolan:2007mj}
%%%%%=====
\begin{align}\label{eq:scalar_ansatz}
  \Phi^{(0)}_{nlm}(t,r,\theta,\varphi) = e^{-i\omega_{nlm}^{(0)}t}e^{im\varphi} R_{nlm}(r) S_{nlm}(\theta),
\end{align}
%%%%% 
where $S_{nlm}(\theta)$ are spheroidal harmonics. 

In the non-relativistic limit, corresponding to $\alpha \equiv M_\mathrm{BH}\mu_\mathrm{s} \ll 1$, the real part $E_{nlm}^{(0)}$ of the eigenfrequencies can be expanded as a power series in $\alpha$~\cite{Baumann:2019eav}:
%%%%%=====
\begin{align}\label{eq:omega}
	E_{nlm}^{(0)} &\approx \mu_\mathrm{s} \Big(1-\frac{\alpha^2}{2\bar{n}^2}-\frac{\alpha^4}{8\bar{n}^4}+\frac{f_{\bar{n}l}}{\bar{n}^3}\alpha^4+\frac{h_l a_* m}{\bar{n}^3}\alpha^5\Big),
\end{align}
%%%%%
with
%%%%%=====
\begin{align}
	f_{\bar{n}l} & \equiv -\frac{6}{2l+1}+\frac{2}{\bar{n}},\\
	h_{l} & \equiv \frac{16}{2l(2l+1)(2l+2)}\,, \textrm{ for } l\geq 1.
\end{align}
%%%%%
Here, $a_* \equiv a/M_\mathrm{BH}$ denotes the dimensionless spin parameter of the BH. In the same limit, the imaginary part $\Gamma_{nlm}^{(0)}$, referred to as the superradiant rate for $\Gamma_{nlm}^{(0)}>0$ and decay rate for $\Gamma_{nlm}^{(0)}<0$, takes the form \cite{Bao:2022hew,Bao:2023xna,Guo:2025dkx}:
%%%%%=====
\begin{align}\label{eq:Gamma}
	\begin{split}
		& \Gamma_{nlm}^{(0)} \approx
    \\
		& -\omega_1 \left(4\kappa\sqrt{M_\mathrm{BH}^2-a^2}\right)^{2l^\prime+1}\frac{\Gamma (n + 2 l^\prime + 2)}{n!}\frac{\sinh (2 \pi  p)}{2 \pi }\times
		\\
		& \frac{\left| \Gamma \left(l^\prime+1-i p+\sqrt{d-p^2}\right) \Gamma \left(l^\prime+1+i p+\sqrt{d-p^2}\right) \right| ^2}{\left[\Gamma ( 2 l^\prime + 1)\Gamma ( 2 l^\prime + 2)\right]^2},
	\end{split}
\end{align}
%%%%%
where $l^\prime \equiv l+\epsilon$, $p \equiv M_\mathrm{BH} r_+ (E_{nlm}^{(0)} - m \Omega_\text{H}) / \sqrt{M_\mathrm{BH}^2-a^2}$, $\kappa \equiv \sqrt{\mu_\mathrm{s}^2-\omega_{0}^2}$, and
%%%%%=====
\begin{subequations}
\begin{align}
	\epsilon & \equiv -\frac{8\alpha^2}{2l+1},
	\\
  \begin{split}
    d & \equiv \frac{8 M_\mathrm{BH} r_+ \omega_{nlm} (r_+\omega_{nlm} - m M_\mathrm{BH} \Omega_\text{H})}{r_+ - r_-} 
    \\
    & \hspace{1cm} - \mu_\mathrm{s}^2 (r_+^2 + a^2) +4 M_\mathrm{BH}^2 (\mu_\mathrm{s}^2 - 3 \omega^2_{nlm}),
  \end{split}
	\\
	\omega_0 & \equiv \mu_\mathrm{s} \sqrt{1 - \frac{2 \alpha^2}{\bar{n}^2 +4\alpha^2 + \bar{n} \sqrt{\bar{n}^2 + 8 \alpha^2}}},
	\\
	\omega_1 & \equiv \frac{\mu_\mathrm{s}^2 - \omega_{0}^2}{\bar{n} \omega_{0} (1 + 4 M_\mathrm{BH}^2 (2\omega_{0}^2 - \mu_\mathrm{s}^2) / \bar{n}^2)}.
\end{align}
\end{subequations}
%%%%%

\section{The environments}
\label{sec:environments}
We consider three different types of environmental effects: galactic halos, equatorial accretion disks and binary companions. In this section, we briefly describe their impact on the background geometry.
\subsection{Galactic Halos}
To describe a BH embedded in a galactic environment within General Relativity, Ref.~\cite{Cardoso:2021wlq} adopts a generalization of Einstein’s construction of a stationary gravitating system, the ``Einstein cluster''. The idea is to model the halo around the BH as an anisotropic fluid with tangential pressure, arising from an ensemble of randomly oriented circular geodesics, with vanishing radial pressure. Under this construction, and assuming spherical symmetry, Ref.~\cite{Cardoso:2021wlq} obtains an exact solution of Einstein’s equations that reduces to the Schwarzschild geometry on small scales and smoothly transitions to a Hernquist-type density distribution \cite{Hernquist:1990be} on large scales, given by
%%%%%=====
\begin{align}\label{eq:metric_galaxy}
    ds^2 = - f_1(r) dt^2 + f_2^{-1}(r) dr^2 + r^2 d\Omega^2\,,
\end{align}
%%%%%
where
%%%%%=====
\begin{subequations}\label{eq:metric_functions}
\begin{align}
    f_1(r) &= \left(1-\frac{2M_\mathrm{BH}}{r}\right)e^\Upsilon,
    \\
    \Upsilon &= -\pi\sqrt{\frac{M_\mathrm{h}}{\xi}} + 2\sqrt{\frac{M_\mathrm{h}}{\xi}}\arctan\frac{r+a_0-M_\mathrm{h}}{\sqrt{M_\mathrm{h}\xi}},
    \\
    \xi &= 2a_0 - M_\mathrm{h} + 4 M_\mathrm{BH},
    \\
    f_2(r) &= 1-\frac{2}{r} \left[M_\mathrm{BH}+\frac{M_\mathrm{h}r^2}{(a_0+r)^2}\left(1-\frac{2M_\mathrm{BH}}{r}\right)^2\right].
\end{align}
\end{subequations}
%%%%%
Here, $M_\mathrm{h}$ and $a_0$ are the total mass and the typical lengthscale of the galactic halo, respectively. For realistic galaxies, these parameters satisfy a clear hierarchy of scales, $M_\mathrm{BH} \ll M_\mathrm{h} \ll a_0$, together with $a_0 \gtrsim 10^4 M_\mathrm{h}$ \cite{Cardoso:2021wlq}. Therefore, we define the perturbative parameter $\epsilon_{\rm h}=M_\mathrm{h} / a_0$.

Writing the Klein-Gordon equation in this background and using the ansatz 
%%%%%=====
\begin{align}
    \Phi = e^{-i\omega t}R(r)Y_{lm}(\theta,\varphi),
\end{align}
%%%%%
where $Y_{lm}(\theta,\varphi)$ are spherical harmonics, we find that $R(r)$ satisfies the following ordinary differential equation:
%%%%%=====
\begin{align}\label{eq:radial_eq_galaxy}
    \begin{split}
        &\frac{\sqrt{f_1(r)f_2(r)}}{r^2}\partial_r\left[r^2\sqrt{f_1(r)f_2(r)}\partial_rR(r)\right] 
        \\
        & \hspace{1cm} + \left[\omega^2 -  f_1(r)\mu_\mathrm{s}^2-\frac{f_1(r)}{r^2}l(l+1)\right] R(r) = 0.
    \end{split}
\end{align}
%%%%%
\subsection{Accretion Disks}\label{sec:Accretion}
The second metric we consider is a Schwarzschild BH-disk model \cite{Kotlarik:2018nbd,Kotlarik:2022spo,Chen:2023akf,Cannizzaro:2024fpz}, a static and axially symmetric spacetime that describes a non-spinning BH surrounded by a thin accretion disk with negligible thickness. The metric is given by \cite{Kotlarik:2022spo,Chen:2023akf,Cannizzaro:2024fpz}:
%%%%%=====
\begin{align}\label{eq:metric_disk}
  \begin{aligned}
    \mathrm{d}s^2  = & -f(r)\mathrm{e}^{2\nu_\mathrm{disk}}\mathrm{d}t^2+\mathrm{e}^{2\lambda_\mathrm{ext}-2\nu_\mathrm{disk}}\frac{\mathrm{d}r^2}{f(r)} 
    \\
    & + r^2\mathrm{e}^{-2\nu_\mathrm{disk}}\left(\mathrm{e}^{2\lambda_\mathrm{ext}}\mathrm{d}\theta^2+\sin^2\theta\mathrm{d}\varphi^2\right),
  \end{aligned}
\end{align}
%%%%% 
where $f(r) \equiv 1 - 2M_\mathrm{BH}/r$, and the disk potential is given by \cite{1963ApJ...138..385T,10.1111/j.1365-2966.2009.14803.x}
%%%%%=====
\begin{align}
  \nu_\mathrm{disk}^{(\mathsf{m},\mathsf{n})}
  =
  -W^{(\mathsf{m},\mathsf{n})} \sum_{k=0}^{\mathsf{m}+\mathsf{n}} \mathcal{Q}_k^{(\mathsf{m},\mathsf{n})} \frac{b^k}{r_b^{k+1}} P_k\left(\frac{|z|+b}{r_b}\right).
\end{align}
%%%%%
Here, the parameter $b$ approximately determines the radial position of the maximum density in the disk. The position and the width of the peak also depend on the coefficients $\mathsf{m}$ and $\mathsf{n}$. The coordinate $r_b$ is expressed in Weyl coordinates as 
\begin{equation}
r_b = \sqrt{\rho^2+(|z|+b)^2} \, ,
\end{equation}
with $\rho^2 = r(r-2M_\mathrm{BH})\sin^2\theta$ and $z = (r-M_\mathrm{BH})\cos\theta$.
Furthermore, $P_j$ denote Legendre polynomials, and the coefficient $\mathcal{Q}_j^{(\mathsf{m},\mathsf{n})}$ is defined as
%%%%%=====
\begin{align}
  \mathcal{Q}_j=
  \begin{cases}
    \sum_{k=0}^\mathsf{n}(-1)^k\binom{\mathsf{n}}{k}\frac{2^{j-k-\mathsf{m}}(2\mathsf{m}+2k-j)!}{(\mathsf{m}+k-j)!(2\mathsf{m}+2k+1)!!} & \hspace{-1mm}\mathrm{if~}j\leq \mathsf{m},
    \\
    \sum_{k=j}^{\mathsf{m}+\mathsf{n}}(-1)^{k-\mathsf{m}}\binom{\mathsf{n}}{k-\mathsf{m}}\frac{2^{j-k}(2k-j)!}{(k-j)!(2k+1)!!} & \hspace{-1mm} \mathrm{if~}j>\mathsf{m}\,,  
  \end{cases}
\end{align}
%%%%%
where we dropped the indices $(\mathsf{m},\mathsf{n})$ to ease the notation.
The normalization factor $W^{(\mathsf{m},\mathsf{n})}$ is defined through
%%%%%=====
\begin{align}
  W^{(\mathsf{m},\mathsf{n})} = (2\mathsf{m}+1)\binom{\mathsf{m}+\mathsf{n}+1/2}{\mathsf{n}}M_\mathrm{d},
\end{align}
%%%%%
ensuring that the total mass of the disk equals $M_\mathrm{d}$.

If the disk is a small perturbation over a Schwarzschild metric, the metric function $\lambda_{\rm ext}$ can be approximated by $\lambda_{\rm ext}\approx \lambda_\mathrm{int}$, where $\lambda_\mathrm{int}$ can be found from the recursive series~\cite{Cannizzaro:2024fpz}:
%%%%%=====
\begin{align}
  \nonumber &\lambda_{\mathrm{int}}^{(0,0)}=-\frac{M_{\mathrm{d}}}{r_{b}}\left(\frac{R_{+}}{b+M_{\mathrm{BH}}}-\frac{R_{-}}{b-M_{\mathrm{BH}}}\right)-\frac{2M_{\mathrm{d}}M_{\mathrm{BH}}}{b^2-M_{\mathrm{BH}}^2}, 
  \\
  \nonumber &\lambda_{\mathrm{int}}^{(0,\mathsf{n}+1)}=\lambda_{\mathrm{int}}^{(0,\mathsf{n})}+\frac{b}{2(\mathsf{n}+1)}\frac{\partial}{\partial b}\lambda_{\mathrm{int}}^{(0,\mathsf{n})},
  \\
  %\begin{split}
  \nonumber
    &\frac{(2\mathsf{m}+1)(2\mathsf{n}+3)}{2\mathsf{m}+2\mathsf{n}+3}\lambda_{\mathrm{int}}^{(\mathsf{m}+1,\mathsf{n})}
    =\lambda_\mathrm{int}^{(\mathsf{m},\mathsf{n})}-b\frac{\partial}{\partial b}\lambda_\mathrm{int}^{(\mathsf{m},\mathsf{n})}\\
    &\hspace{4.5 cm}+\frac{4\mathsf{m}(\mathsf{n}+1)}{2\mathsf{m}+2\mathsf{n}+3}\lambda_\mathrm{int}^{(\mathsf{m},\mathsf{n}+1)},
  %\end{split}
\end{align}
%%%%%
with $R_{\pm}=\sqrt{\rho^2+(|z|\mp M_\mathrm{BH})^2}$. For simplicity, in the following section we fix $\mathsf{m}$ and $\mathsf{n}$ to the characteristic values $0$ and $1$, respectively. 

Because the metric functions $\nu_{\rm disk}(r,\theta)$ and $\lambda_{\rm ext}(r,\theta)$ have nonseparable radial and angular dependence, the Klein–Gordon equation is not separable.

A naive decomposition in spherical harmonics would therefore induce couplings between different multipole $l$-modes. This issue can be circumvented by employing the projection method~\cite{Cano:2020cao,Chen:2022ynz}, where the wave equation is projected onto a basis of associated Legendre functions and the orthogonality relations are used to isolate the diagonal contributions at linear order in the deformation parameter $\epsilon_{\rm d}=M_{\rm d}/M_{\rm BH}$. In this way, the coupled system reduces to an effective Schr\"odinger-like radial equation for each mode:
%%%%%=====
\begin{align}\label{eq:radial_eq}
  \begin{split}
    &\left[1+\epsilon_{\rm d}\left(2 \mathcal{V}_j - 2 \mathcal{L}_j\right)b_{lm}^{j}\right] \frac{f}{r^{2}}\partial_{r}\left[r^{2}f\partial_{r}R(r)\right]
    \\
    &+\left[(1-2\epsilon_{\rm d}\mathcal{V}_j b_{lm}^{j})\omega^{2} - f \mu_\mathrm{s}^2 \right]R(r)
    \\
    &-\left[\frac{l(l+1)f}{r^{2}}+
    \frac{\epsilon_{\rm d}  f}{r^{2}} U_{lm}(r)\right]R(r) =0,
  \end{split}
\end{align}
where we defined $U_{lm}(r)  =2 \mathcal{V}_j a_{lm}^{j}  -\left(2 \mathcal{V}_j - 2 \mathcal{L}_j\right)c_{lm}^{j}$, the functions $\mathcal{V}_j(r)$ and $\mathcal{L}_j(r)$ describe the series coefficients of $\nu_{\rm disk}$ and $\lambda_{\rm int}$ after expanding the metric functions in powers of $x=\cos\theta$, while coefficients $a_{lm}^j$, $b_{lm}^j$, and $c_{lm}^j$ arise from projecting the Klein-Gordon equation onto the basis of associated Legendre functions $P_l^m(x)$. The explicit form of the projection and series coefficients is reported in Appendix~\ref{appendix:projection_method}.

\subsection{Binary companions}\label{subsec:binarycompanion}
Finally, we consider tidal perturbations induced by binary companions. For this case, we consider a Newtonian approximation in which the tidal potential $\delta V$ takes the form \cite{Tomaselli:2023ysb}
%%%%%=====
\begin{equation}\label{eq:delta_V_companions}
  \delta V = -\sum^{\infty}_{l_*=0} \sum^{l_*}_{m_*=-l_*} \frac{4\pi q M_\mathrm{BH}}{2l_*+1} Y_{l_*m_*}(\theta^A_*)Y^*_{l_*m_*}(\theta^A)F(r),
\end{equation}
%%%%%
where
%%%%%=====
\begin{align}
  F(r) = 
  \begin{cases}
    \frac{r^{l_*}}{R_*^{l_*+1}} \Theta(R_*-r)+ \frac{R_*^{l_*}}{r^{l_*+1}} \Theta(r-R_*) &\text{for }l_*\not=1,
    \\
    \left(\frac{R_*}{r^2}-\frac{r}{R_*^2}\right) \Theta(r-R_*) &\text{for }l_*=1\,.
  \end{cases}
\end{align}
%%%%%
and $\Theta$ denotes the Heaviside step function, $q \equiv M_*/M_\mathrm{BH}$ is the mass ratio between the companion and the BH, $\theta^A=(\theta,\varphi)$, and $(R_*,\theta_*,\varphi_*)$ specifies the position of the companion.

The multipole expansion of the companion's gravitational potential is valid as long as the binary separation $R_*$ exceeds the characteristic size of the cloud $r_\mathrm{c}\sim M_{\rm BH}(\bar{n}/\alpha)^2$, i.e., $R_* \gg r_\mathrm{c}$, or the mass ratio is sufficiently small $q\ll 1$. More succinctly, we require $q\,r_\mathrm{c}/\max(R_*,r_\mathrm{c})\ll 1$~\cite{Roy:2025qaa}. Furthermore, on the dynamical timescale of the scalar cloud, of order $\mu_\mathrm{s}^{-1}$, we assume that the position of the companion remains fixed, i.e., $\mu_\mathrm{s} \gg \Omega_*$. Here, $\Omega_*$ denotes the orbital angular velocity, which for a circular orbit is given by
%%%%%=====
\begin{align}
  \Omega_{*} = \sqrt{\frac{M_\mathrm{BH}(1+q)}{R_*^3}}.
\end{align}
%%%%%
The adiabatic condition $\mu_\mathrm{s} \gg \Omega_*$ can equivalently be written as $R_* \gg r_\mathrm{c}^{1/3}\bar{n}^{-2/3}M_\mathrm{BH}^{2/3}(1+q)^{1/3}$. In the regime $\alpha<1$ and $R_*\gg r_\mathrm{c}$, this condition is always satisfied, as long as $q$ is not much larger than unity.\footnote{For $R_*\gg r_c$ and $q\lesssim 1$ mass transfer can also be neglected~\cite{Baumann:2018vus}.}  

Finally, we note that in this case a complete relativistic description of the BH-binary spacetime is not available. Constructing the full geometry would require a metric reconstruction of the binary system, which is considerably more involved. We therefore model the companion only as an external perturbation through its Newtonian potential, restricting our calculation to the regime $R_* \gtrsim r_\mathrm{c}$, to ensure that our assumptions remain valid and that the Newtonian potential provides a reasonably good approximation for the binary-companion perturbation.

\section{First-order shifts}
\label{sec:perturbation_theory}
The environments presented above induce only mild modifications to the background geometry for realistic astrophysical parameters. Their impact on the scalar field eigenfrequencies discussed in Sec.~\ref{sec:scalar_superradiance} can therefore be treated perturbatively. In the following, we present different methods to compute first-order shifts of these eigenfrequencies due to small metric perturbations of a Kerr spacetime. One method is based on a non-relativistic approximation and two methods use a fully relativistic approach. We then assess their accuracy through a comparison with fully numerical results.

\subsection{Non-relativistic perturbation method}
At leading-order in the non-relativistic limit $\alpha\ll1$, the Klein-Gordon equation~\eqref{eq:KGeq} reduces to the Schrödinger equation \cite{Baumann:2018vus}
%%%%%=====
\begin{align}
    i \frac{\partial}{\partial t} \psi^{(0)}(t,r,\theta,\varphi) = \left[ -\frac{1}{2\mu_\mathrm{s}}\nabla^2 -\frac{\alpha}{r} \right] \psi^{(0)}(t,r,\theta,\varphi),
\end{align}
%%%%%
where $\Phi^{(0)}=\psi^{(0)} e^{-i\mu_\mathrm{s} t} / \sqrt{\mu_\mathrm{s}}$, and $\psi^{(0)}$ is assumed to vary on a timescale that is much longer than $\mu^{-1}$. 

In this limit, the BH is substituted by a point particle with mass $M_{\rm BH}$ and the dissipative boundary conditions at the outer horizon are substituted by regular boundary conditions at the origin, leading to the same solutions as the hydrogen atom bound states~\cite{Brito:2015oca}:
%%%%%=====
\begin{align}
    \psi^{(0)}_{nlm}(t,r,\theta,\varphi) = e^{-i(E^{(0)}_{nlm}-\mu_\mathrm{s})t}R_{nl}^\mathrm{H}(r)Y_{lm}(\theta,\varphi),
\end{align}
%%%%%
with
%%%%%=====
\begin{align}
    R_{nl}^\mathrm{H}(r)=\sqrt{\left(\frac{2}{\bar{n}r_\mathrm{B}}\right)^3 \frac{n!}{2\bar{n}(\bar{n}+l)!}}\tilde{r}^l e^{-\tilde{r}/2}L_{n}^{2l+1}(\tilde{r}),
\end{align}
%%%%%
where $\tilde{r}\equiv 2r/(\bar{n}r_\mathrm{B})$, $r_\mathrm{B}=1/(\mu_\mathrm{s}\alpha)$ is the Bohr radius, and $L_{n}^{2l+1}$ are associated Laguerre polynomials. Note that $\psi^{(0)}_{nlm}$ has been normalized such that $\langle \psi^{(0)}_{nlm}|\psi^{(0)}_{nlm}\rangle=1$, where $\langle \cdot|\cdot\rangle$ denotes the usual $L^2$ space inner product. 

If the system is perturbed by a Newtonian potential
$\delta V=\delta V^{(1)}+\delta V^{(2)}+\cdots$, the eigenfrequencies and corresponding eigenfunctions admit the perturbative expansions
$\omega_{nlm}=\omega_{nlm}^{(0)}+\delta\omega_{nlm}^{(1)}+\delta\omega_{nlm}^{(2)}+\cdots$ and
$\psi_{nlm}=\psi_{nlm}^{(0)}+\delta\psi_{nlm}^{(1)}+\delta\psi_{nlm}^{(2)}+\cdots$. 
The first-order eigenfrequency shift can then be computed using standard quantum mechanical perturbation theory and is given by
%%%%%=====
\begin{align}
    \delta \omega_{nlm}^{(1)} = \langle\psi^{(0)}_{nlm}|\mu_\mathrm{s} \delta V^{(1)}|\psi^{(0)}_{nlm}\rangle\,.
\end{align}
%%%%%
Since dissipation at the event horizon is neglected in the Schrödinger treatment, this expression only captures the correction to the real part of the frequency.

\subsection{Relativistic perturbation method}

The non-relativistic perturbation method is valid only in the regime $\alpha \ll 1$. More importantly, the physical dissipative boundary condition at the BH horizon renders the system non-Hermitian, so that the quasibound modes fail to satisfy orthogonality under the standard $L^2$ space inner product. Therefore, the non-relativistic perturbation method is not suitable for computing shifts in the decay or superradiant rates.

Recently, Ref.~\cite{Cannizzaro:2023jle} extended the gravitational bilinear form of Ref.~\cite{Green:2022htq} to massive scalar fields around Kerr BHs, constructing a conserved product based on the spacetime symmetries. Crucially, quasibound states are orthogonal with respect to this bilinear form, thereby providing the foundation for relativistic first-order perturbation theory (see also Ref.~\cite{Cannizzaro:2025vpb} for an extension to quasinormal modes). We stress, however, that orthogonality with respect to this bilinear form does not by itself imply a completeness relation, a point that becomes crucial at second order, as we discuss in the next section.

The bilinear form is defined as~\cite{Cannizzaro:2023jle}
%%%%%=====
\begin{align}
  \langle\langle\Phi_1^{(0)},\Phi_2^{(0)}\rangle\rangle = \Pi_\Sigma[\mathcal{J}\Phi_1^{(0)},\Phi_2^{(0)}],
\end{align}
%%%%%
where $\mathcal{J}$ denotes the $t$-$\varphi$ symmetry operator, whose action on a scalar field corresponds to the transformation $t \rightarrow -t$ and $\varphi \rightarrow -\varphi$, and 
%%%%%=====
\begin{align}
  \Pi_\Sigma[\Phi_1^{(0)},\Phi_2^{(0)}] \equiv \int_\Sigma (\Phi_1^{(0)}\nabla_\mu^{(0)}\Phi_2^{(0)}-\Phi_2^{(0)}\nabla_\mu^{(0)}\Phi_1^{(0)})n^\mu d\Sigma.
\end{align}
%%%%%
Here, $\Sigma$ denotes a spacelike hypersurface on a Kerr geometry, and $n^\mu$ is the corresponding unit normal vector. The indices $1,2$ refer to two, possibly distinct, quasibound state solutions. In Boyer-Lindquist coordinates, the bilinear form can be rewritten as \cite{Cannizzaro:2023jle}
%%%%%=====
\begin{align}\label{eq:bilinear_form}
    \begin{split}
        \langle\langle \Phi_1^{(0)}, \Phi_2^{(0)} \rangle\rangle = & \int_{r_+}^{\infty} \mathrm{d}r \int_{S^2} \mathrm{d}^2\Omega \times \\
        &\Bigg[ \frac{2M_{\mathrm{BH}}ra}{\Delta} (\mathcal{J}\Phi_1^{(0)} \partial_\varphi \Phi_2^{(0)} - \Phi_2^{(0)} \partial_\varphi \mathcal{J}\Phi_1^{(0)}) \\
        & + \frac{\Sigma}{\Delta} \left( r^2 + a^2 + \frac{2M_{\mathrm{BH}}ra^2}{\Sigma} \sin^2 \theta \right) \\
        & \times (\mathcal{J}\Phi_1^{(0)} \partial_t \Phi_2^{(0)} - \Phi_2^{(0)} \partial_t \mathcal{J}\Phi_1^{(0)}) \Bigg]\,,
    \end{split}
\end{align}
%%%%%
where $\mathrm{d}^2\Omega=\sin\theta \mathrm{d}\theta \mathrm{d}\varphi$. 

Because the radial relativistic functions $R_{nlm}(r)$ [see Eq.~\eqref{eq:scalar_ansatz}] diverge at the horizon as $r\rightarrow r_+$ whenever $\Gamma^{(0)}<0$, to ensure the convergence of the radial integral appearing in the bilinear form, we deform the integration path into a complex contour $\mathcal{C}$. Defining the tortoise coordinate $r_*$ through 
%%%%%=====
\begin{align}\label{eq:tortoise}
    \frac{dr}{dr_*} = \frac{\Delta}{r^2+a^2}\,,
\end{align}
%%%%%
the contour is chosen such that \cite{Green:2022htq,Cannizzaro:2023jle}
%%%%%=====
\begin{align}\label{eq:arg_contour}
    \arg r_* + \arg\left[k_{{\rm H},1}^{(0)}+k_{{\rm H},2}^{(0)}\right] = -\pi/2, \quad r_*\rightarrow-\infty\,,
\end{align}
%%%%%
while running along the real axis elsewhere. Here, we defined $k_{{\rm H},1/2}^{(0)}=\omega_{1/2}^{(0)}-m_{1/2}\Omega_\mathrm{H}$. With this contour prescription, the explicit form of Eq.~\eqref{eq:bilinear_form} in the Kerr background becomes \cite{Cannizzaro:2023jle}
%%%%%=====
\begin{align}\label{eq:bilinear_kerr}
    \begin{split}
        &\langle\langle \Phi_{1}^{(0)} , \Phi_{2}^{(0)} \rangle \rangle = 2\pi i\delta_{m_{1}m_{2}}e^{i(\omega_{1}^{(0)}-\omega_{2}^{(0)})t}\int\limits_{\mathcal{C}}dr\frac{K}{\Delta}R_{1}R_{2},
    \end{split}
\end{align}
where
\begin{align}
    \begin{split}
        K = & -\left(\omega_{1}^{(0)}+\omega_{2}^{(0)}\right)\left[\left(r^{2}+a^{2}\right)^{2}I_{\theta,1}-\Delta a^{2}I_{\theta,3}\right]
        \\
        & + 2M_\mathrm{BH}ra\left(m_{1}+m_{2}\right)I_{\theta,1},
    \end{split}
    \\
    I_{\theta,1}=&\int_{0}^{\pi}S_{1}(\theta)S_{2}(\theta)\sin\theta d\theta,
    \\
    I_{\theta,3}=&\int_{0}^{\pi}S_{1}(\theta)S_{2}(\theta)\sin^{3}\theta d\theta.
\end{align}
%%%%%

An equivalent and technically simpler regularization procedure can be obtained by performing the radial integration directly along the real axis and subtracting the divergent contribution {\it a posteriori}, following the counter-term subtraction method introduced in Refs.~\cite{Cannizzaro:2023jle,Sberna:2021eui}. While originally developed for a Schwarzschild spacetime, in this work we extend the method to the Kerr case. We refer to Appendix~\ref{appendix:counterterm} for the detailed derivation and report below the final regularized form of the bilinear product:
%%%%%=====
\begin{align}\label{eq:regular_bilinearform}
    \begin{split}
        &\langle\langle \Phi_{1}^{(0)} , \Phi_{2}^{(0)} \rangle \rangle = \ 2\pi i\delta_{m_{1}m_{2}}e^{i(\omega_{1}^{(0)}-\omega_{2}^{(0)})t}\lim_{\bar{r}\rightarrow r_+}\Bigg[ 
        \\
        &\int_{\bar{r}}^\infty dr \frac{K}{\Delta} R_1 R_2 + \frac{K(\bar{r})}{\bar{r}^2+a^2} \frac{iR_1(\bar{r}) R_2(\bar{r})}{\omega_{1}^{(0)}+\omega_{2}^{(0)} - (m_1+m_2) \Omega_\mathrm{H}}\Bigg]\,,
    \end{split}
\end{align}
%%%%%
which reduces to the one obtained in~\cite{Cannizzaro:2023jle} in the non-spinning limit.  

Based on the bilinear form, eigenfrequency shifts can be computed following the same approach as in quantum mechanical perturbation theory. As discussed in Ref.~\cite{Cannizzaro:2023jle}, the calculation can be organized in either a Hamiltonian formulation or in a covariant formulation. In the former, and within a semi-Newtonian approximation in which metric perturbations are encoded through an effective potential $\delta V^{(1)}$, the first-order frequency shift reads \cite{Cannizzaro:2023jle}
%%%%%=====
\begin{align}\label{eq:frequency_shift}
  \delta \omega_{nlm}^{(1)} = \omega_{nlm}^{(0)} \frac{\langle\langle \Phi^{(0)}_{nlm},\delta V^{(1)} \Phi^{(0)}_{nlm}\rangle\rangle}{\langle\langle \Phi^{(0)}_{nlm},\Phi^{(0)}_{nlm}\rangle\rangle}.
\end{align}
%%%%%
In the covariant formulation, the Klein-Gordon operator is instead directly expanded as $\mathcal O=\mathcal{O}^{(0)} + \delta\mathcal O^{(1)} + \delta\mathcal O^{(2)} + \cdots$, with  $\mathcal{O}^{(0)}\equiv -\nabla^{(0)\mu} \nabla^{(0)}_{\mu} + \mu_\mathrm{s}^2$, and $\delta\mathcal O^{(1)}$, $\delta\mathcal O^{(2)}$ denoting first and second-order terms in the expansion parameter, respectively. The corresponding first-order shift to the eigenfrequencies is \cite{Cannizzaro:2023jle}
%%%%%=====
\begin{align}\label{eq:frequency_shift_cov}
  -i\delta\omega_{nlm}^{(1)} = \frac{(\delta\mathcal{O}^{(1)})_{nlm,nlm}}{\langle\langle\Phi^{(0)}_{nlm},\Phi^{(0)}_{nlm}\rangle\rangle}.
\end{align}
%%%%%

In a Schwarzschild background, one can use an expansion in spherical harmonics, and $\mathcal{O}^{(0)}$ is given by
%%%%%=====
\begin{align}
  f(r)\mathcal{O}^{(0)} &= -f(r)^2\partial_{r}^2-\frac{f(r)}{r^{2}}\partial_{r}\left[r^{2}f(r)\right]\partial_{r}\nonumber
  \\
  & \hspace{0.4cm} +\left[\partial_t^2+f(r)\mu_{\mathrm{s}}^{2}+\frac{f(r)}{r^{2}}l(l+1)\right],
\end{align}
%%%%%
and $(\delta\mathcal{O})_{nlm,nlm}^{(1)}$ can be written as
%%%%%=====
\begin{align}
  (\delta\mathcal{O}^{(1)})_{nlm,nlm} ={-i} \frac{\langle\langle\Phi^{(0)}_{nlm},f(r)\delta\mathcal{O}^{(1)}\Phi^{(0)}_{nlm}\rangle\rangle}{2\omega_{nlm}^{(0)}}.
\end{align}
%%%%%

In the following, when a time-domain operator is evaluated on a mode with frequency $\omega$, we use the replacement $\partial_t^2\rightarrow-\omega^2$ to obtain its frequency-domain form.

\subsection{Results} \label{sec:galaxy}

%%%%%FFFFF
\begin{figure*}[htb]
  \includegraphics[width=0.9\linewidth]{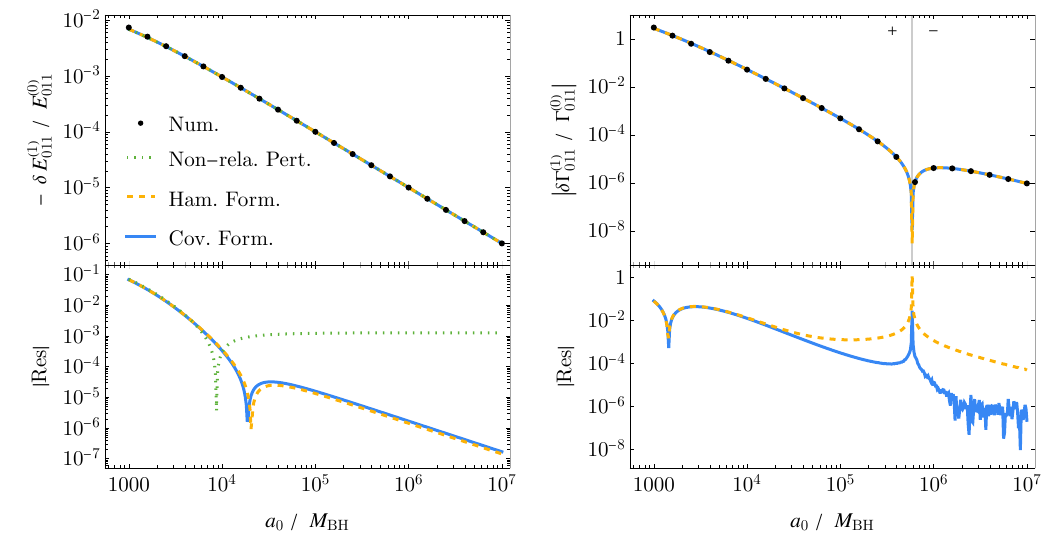}
  \caption{First-order relative eigenfrequency (top panels) shifts for the $(n,l,m)=(0,1,1)$ mode, induced by the galactic environment model, and the corresponding absolute fractional residuals (bottom panels) as functions of $a_0/M_\mathrm{BH}$ (as defined by Eq.~\eqref{eq:residual}). The zeroth-order eigenfrequencies $\omega_{nlm}^{(0)}\equiv E_{nlm}^{(0)}+i\Gamma^{(0)}_{nlm}$ are computed in a Schwarzschild background. Black points denote $(\omega_{nlm}-\omega_{nlm}^{(0)})/\omega_{nlm}^{(0)}$, where $\omega_{nlm}$ is the non-perturbative numerical solution obtained by directly integrating Eq.~\eqref{eq:radial_eq_galaxy}. The blue solid, orange dashed, and green dotted lines represent the results from the covariant formulation, the Hamiltonian formulation, and the non-relativistic perturbation method, respectively. The left panels show the relative corrections and their corresponding absolute fractional residuals for the energy levels, while the right panels display those for the decay rates. In the top right panel, the vertical line marks a zero-crossing of $\delta\Gamma_{011}^{(1)}/\Gamma_{011}^{(0)}$ and values to the left (right) of this line are positive (negative). The other parameters are fixed at $M_\mathrm{h} = 10 M_\mathrm{BH}$ and $\alpha = 0.1$.}
  \label{fig:diff_a0}
\end{figure*}
%%%%%

%%%%%FFFFF
\begin{figure*}[htb]
  \includegraphics[width=0.9\linewidth]{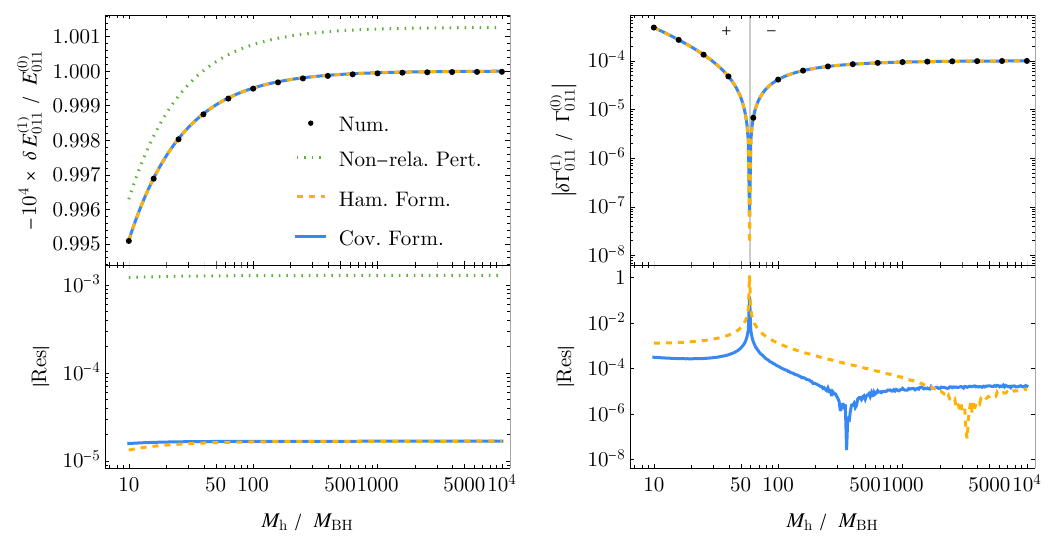}
  \caption{Same as Fig.~\ref{fig:diff_a0}, but now showing results as a function of $M_\mathrm{h}/M_\mathrm{BH}$ and fixing the other parameters to $a_0 = 10^4 M_\mathrm{h}$ and $\alpha = 0.1$.}
  \label{fig:diff_Mh}
\end{figure*}
%%%%%

%%%%%FFFFF
\begin{figure*}[htb]
  \includegraphics[width=0.9\linewidth]{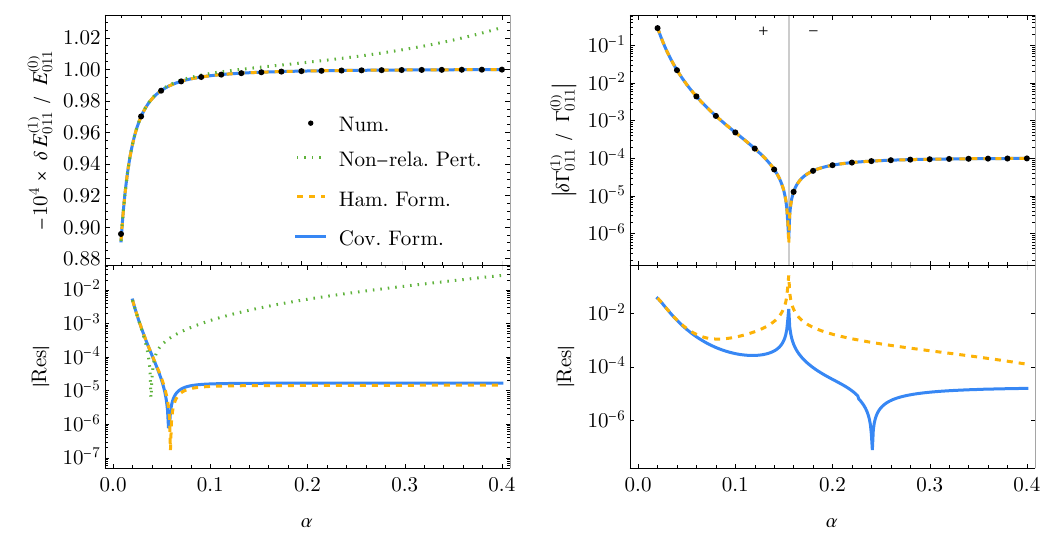}
  \caption{Same as Fig.~\ref{fig:diff_a0}, but now showing results as a function of $\alpha$ and fixing the other parameters to $a_0 = 10^4 M_\mathrm{h}$ and $M_\mathrm{h} = 10M_\mathrm{BH}$.  }
  \label{fig:diff_alpha}
\end{figure*}
%%%%%

We now use the formalism laid out above to compute first-order frequency shifts due to the different environments discussed in Sec.~\ref{sec:environments} and, when possible, compare the perturbative predictions with non-perturbative results for the eigenfrequencies obtained by numerically solving the relevant equations in the full metric background. This allows us to assess the validity of the different first-order approaches (i.e. Hamiltonian versus covariant formulations)  and identify their regimes of applicability. 

Let us start by considering the galactic halo environment, for which the perturbing potential and operator are given by 
%%%%%=====
\begin{align}
    \delta V^{(1)} = -\frac{M_\mathrm{h}}{a_0+r},
\end{align}
%%%%%
and 
%%%%%=====
\begin{align}\label{eq:Operator_1storder_halo}
    \begin{split}
        f(r)\delta\mathcal{O}^{(1)} &= \frac{2 M_\mathrm{h} (r-2M_\mathrm{BH})}{(a_0+r)^2} f(r)^2\partial_{r}^2
        \\
        &+\frac{M_\mathrm{h} (r-2M_\mathrm{BH}) \left(2 a_0 (r-M_\mathrm{BH})+r^2\right)}{(r-M_\mathrm{BH}) (a_0+r)^3}\\ 
        &\times \frac{f(r)}{r^{2}}\partial_{r}\left[r^{2}f(r)\right]\partial_{r}
         +\frac{2 M_\mathrm{h} }{a_0+r} \partial_t^2\,,
    \end{split}
\end{align}
%%%%%
respectively. The spectrum depends on three main parameters: the halo lengthscale $a_0$, the halo mass $M_{\rm h}$, and the mass coupling $\alpha$. In Figs.~\ref{fig:diff_a0}--\ref{fig:diff_alpha}, we show how the energy levels and decay rates vary as these parameters are changed. For the non-perturbative numerical results, we define the frequency shift as $\omega_{nlm}-\omega_{nlm}^{(0)}$, where $\omega_{nlm}$ is obtained by numerically solving Eq.~\eqref{eq:radial_eq_galaxy} with appropriate boundary conditions, i.e. ingoing waves at the BH horizon and exponentially decaying solutions at infinity. Although this quantity also contains higher-order contributions, these should remain negligible within the perturbative regime considered here. Overall, the shifts remain small across the parameter space explored, confirming that realistic galactic halos induce only mild modifications to the spectrum. As expected, lower halo compactness $M_{\rm h}/a_0$ corresponds to weaker shifts in both the real and imaginary parts of the eigenfrequency. Interestingly, the value of $a_0$ at which the decay-rate correction changes sign is the same in Figs.~\ref{fig:diff_a0} and~\ref{fig:diff_Mh}. 
Together with Fig.~\ref{fig:diff_alpha}, this shows that the zero-crossing condition for $\delta\Gamma_{011}^{(1)}$ depends on $a_0$ and $\alpha$ but is independent of $M_\mathrm{h}$.
To confirm that this behavior persists in the rotating case, we further employ the perturbing potential $\delta V$ together with the bilinear form defined in the Kerr background. We find that the inclusion of BH spin does not qualitatively modify this picture.

Let us now compare the perturbative shifts with the non-perturbative numerical solutions. 
To quantify the agreement between the perturbative approach and non-perturbative numerical results, we define the first-order fractional residual as
%%%%%=====
\begin{align}\label{eq:residual}
    {\rm Res \equiv \frac{1st\ order}{Num.} - 1}.
\end{align}
%%%%%
The first feature one can immediately appreciate is the limitation of the non-relativistic approach. In particular, the latter predicts only a correction to the real part of the spectrum, with no shift in the imaginary part, as discussed above. Moreover, the non-relativistic approximation remains accurate only in the small-$\alpha$ regime (see Fig.~\ref{fig:diff_alpha}). Even in this limit, however, the relativistic approaches provide a more accurate description of the spectrum, as can be directly seen from the residuals shown in the lower panels of Figs.~\ref{fig:diff_a0}--\ref{fig:diff_alpha}. 

We can also directly compare the Hamiltonian and covariant formulations to compute the eigenfrequency shifts. Both formulations accurately reproduce the non-perturbative numerical shifts across the parameter space explored, including the imaginary part of the spectrum. In general, the covariant formulation gives a more precise agreement with the numerical solutions than the Hamiltonian approach, as visible from the residuals of Figs.~\ref{fig:diff_a0}--\ref{fig:diff_alpha}. This is expected since the identification of the Hamiltonian with the gravitational potential requires an additional semi-Newtonian approximation, compared with the covariant framework which includes all relativistic corrections up to first order in the perturbation parameter~\cite{Cannizzaro:2023jle}.

%%%%%FFFFF
\begin{figure*}[htb]
  \includegraphics[width=0.9\linewidth]{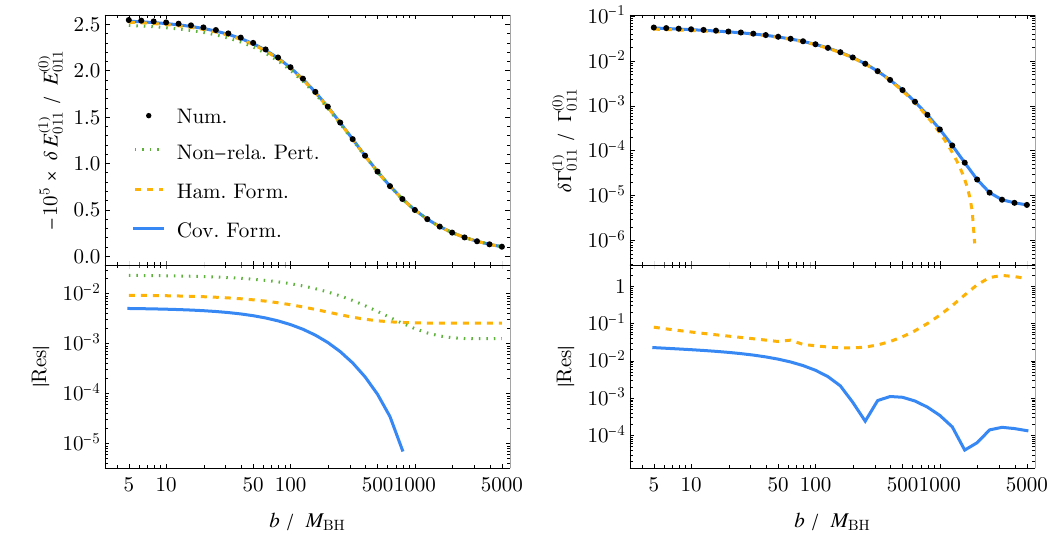}
  \caption{First-order relative eigenfrequency (top panels) shifts for the $(n,l,m)=(0,1,1)$ mode, induced by the disk environment model, and the corresponding absolute fractional residuals (bottom panels) as functions of $b/M_\mathrm{BH}$. The zeroth-order eigenfrequencies $\omega_{nlm}^{(0)}\equiv E_{nlm}^{(0)}+i\Gamma^{(0)}_{nlm}$ are again computed in a Schwarzschild background. Black points denote $(\omega_{nlm}-\omega_{nlm}^{(0)})/\omega_{nlm}^{(0)}$, where $\omega_{nlm}$ is the non-perturbative numerical eigenfrequency obtained by numerically integrating Eq.~\eqref{eq:radial_eq}. The blue solid, orange dashed, and green dotted lines show the results obtained with the covariant formulation, the Hamiltonian formulation, and the non-relativistic perturbation method, respectively. The left panels show the relative corrections and their corresponding absolute fractional residuals for the energy levels, while the right panels display those for the decay rates. The other parameters are fixed at $\epsilon_{\rm d} = 0.01$ and $\alpha = 0.1$.}
  \label{fig:diff_b}
\end{figure*}
%%%%%

We can now turn to the disk case, checking whether the conclusions obtained with galactic halos hold for an environment with a different morphology. To obtain the numerical non-perturbative results, we numerically integrate Eq.~\eqref{eq:radial_eq}, truncating at $j_{\rm max}$ chosen so that the spectrum converges (see Appendix~\ref{sec:disk_shifts} for a convergence analysis). On the other hand, to compute the frequency shifts using the perturbative formalisms, we define the operators: 
$\delta V =\nu_\mathrm{disk}$ and
%%%%%=====
\begin{align}
  \begin{split}
    f(r)\delta\mathcal{O}^{(1)} = \epsilon_{\rm d} \Bigg\{-\left(2 \mathcal{V}_j - 2 \mathcal{L}_j\right) \frac{b_{lm}^{j}}{r^{2}}f(r)\partial_{r}\left[r^{2}f(r)\partial_{r}\right] 
    \\
    + \left[-2\mathcal{V}_j b_{lm}^{j} \partial_t^{2} 
    + \frac{f(r)U_{lm}(r)}{r^{2}} \right] \Bigg\}.
  \end{split}
\end{align}
%%%%%

Overall, the qualitative behavior is very similar to the halo case. In particular, the environmental corrections remain perturbatively small for realistic disk parameters, the non-relativistic approximation has the same limitations and the covariant formalism remains the most accurate one. For this reason, we move the corresponding plots and a more detailed discussion to Appendix~\ref{sec:disk_shifts}, and focus here only on a crucial difference specific to accretion disks. 
While the Hamiltonian approach remains reasonably accurate in the analysis we did with the galactic halo environment, in the disk case it exhibits regions of parameter space where it fails to correctly reproduce the shifts to the decay rates. This is shown in Fig.~\ref{fig:diff_b}, where we compare the non-perturbative numerical results against the perturbative frequency shifts as a function of the parameter $b/M_{\rm BH}$. In the large-$b/M_{\rm BH}$ limit, the Hamiltonian approach reproduces well the energy shifts, yet it fails to completely reproduce the decay-rate shifts.
This point is particularly important in view of realistic BH binaries, that we discuss below. In such systems, a fully relativistic treatment based on the covariant formalism is not readily available, since no exact stationary spacetime describing the binary geometry is known. Constructing the corresponding metric would require reconstructing the binary metric, which is considerably more involved. As a result, the most direct approach consists in modeling the companion through an effective gravitational potential within the Hamiltonian framework. It is therefore crucial to understand the regime of validity of this approximation and identify the situations in which it may fail.

\subsection{On the validity of the Hamiltonian framework}

The origin of the failure of the Hamiltonian framework to correctly capture the decay-rate shifts in the disk case, can be understood directly by inspecting the metric around the peak of the scalar field's eigenfunctions, $r\sim r_\mathrm{c}=\bar{n}^2 M_{\rm BH}/\alpha^2$. In the Hamiltonian formulation, the metric perturbation is encoded only through the semi-Newtonian potential $\delta V=\nu_\mathrm{disk}$, which captures perturbations to the lapse function (see~\cite{Cannizzaro:2023jle}). For the $(\mathsf{m},\mathsf{n})=(0,1)$ disk that we consider, expanding in powers of $r/b$ in the region $r\sim r_\mathrm{c}$ and taking $b\gg r_\mathrm{c}$ gives
%%%%%=====
\begin{align}
  \nu_\mathrm{disk} &\simeq \nu_0\equiv -\frac{M_\mathrm{d}}{2b},
  \\
  \lambda_\mathrm{ext} &= \mathcal{O}\!\left(\nu_0\frac{M_\mathrm{BH}r}{b^2}\right)
  \ll \nu_0 .
\end{align}
%%%%%
Thus, the perturbation to the Schwarzschild metric due to the disk  is nearly constant close to the peak of the scalar field's eigenfunctions and the leading-order BH-disk metric in this region is given by
%%%%%=====
\begin{align}
  ds^2\simeq -e^{2\nu_0}fdt^2
  +e^{-2\nu_0}\left(\frac{dr^2}{f}+r^2d\Omega^2\right)\,.
\end{align}
%%%%%
This metric contains both a time-redshift correction and a spatial rescaling with respect to the Schwarzschild metric written in Schwarzschild coordinates. Changing coordinates to $t'=e^{\nu_0}t$ and $r'=e^{-\nu_0}r$, the spatial rescaling shifts the Schwarzschild mass scale to $M_\mathrm{BH}\rightarrow e^{-\nu_0}M_\mathrm{BH}$, and hence shifts $\alpha=M_\mathrm{BH}\mu_\mathrm{s}$ in a similar way. Since the Hamiltonian formulation only takes into account corrections to the lapse function, it captures the time-redshift contribution, schematically $\delta\omega/\omega\sim \nu_0$, but misses the spatial-rescaling correction. Since $\delta\alpha/\alpha=-\nu_0+\mathcal{O}(\nu_0^2)$, Eq.~\eqref{eq:omega} gives
%%%%%=====
\begin{align}
  \left(\frac{\delta E}{E}\right)_\mathrm{sp}
  \simeq
  \frac{1}{E_{nlm}^{(0)}}
  \frac{\partial E_{nlm}^{(0)}}{\partial\alpha}
  \delta\alpha
  =
  \frac{\alpha^2}{\bar n^2}\nu_0
  +\mathcal{O}(\nu_0\alpha^4),
\end{align}
%%%%%
where the subscript ``sp'' denotes the contribution induced by the spatial rescaling.
The missed contribution to the real part is therefore suppressed by $\alpha^2/\bar n^2$ relative to the time-redshift contribution. For the imaginary part, the leading small-$\alpha$ scaling is
%%%%%=====
\begin{align}
  \Gamma_{nlm}^{(0)}\propto \alpha^{4l+5}.
\end{align}
%%%%%
The same spatial rescaling gives
%%%%%=====
\begin{align}
  \left(\frac{\delta\Gamma}{\Gamma}\right)_\mathrm{sp}
  \simeq
  \frac{1}{\Gamma_{nlm}^{(0)}}
  \frac{\partial \Gamma_{nlm}^{(0)}}{\partial\alpha}
  \delta\alpha
  =
  -(4l+5)\nu_0 .
\end{align}
%%%%%
For the $l=1$ mode, the omitted contribution is therefore $\left(\delta\Gamma/\Gamma\right)_\mathrm{sp}\simeq -9\nu_0$, enhanced by a factor $9$ rather than suppressed by $\alpha^2/\bar n^2$. This explains why the Hamiltonian approximation captures the energy shifts but fails to fully capture the decay-rate shifts in the large-$b/M_{\rm BH}$ disk regime.

This interpretation also explains why no analogous discrepancy appears for $b\lesssim r_\mathrm{c}$. In this regime the disk perturbations cannot be approximated as being a simple constant correction over the whole cloud, and therefore the spatial components proportional to $\nu_\mathrm{disk}$ cannot be interpreted as inducing a simple rescaling of the Schwarzschild mass scale. Although the full covariant operator still contains these spatial-metric contributions, they multiply spatial derivatives of the eigenfunction which are suppressed relative to the leading lapse or rest-mass coupling retained in the potential $\delta V=\nu_\mathrm{disk}$, by $M_\mathrm{BH}/r_\mathrm{c}\sim\alpha^2/\bar n^2$. They consequently do not generate the enhanced $(4l+5)\nu_0$ contribution to $\delta\Gamma/\Gamma$ that appears in the constant-potential large-$b$ limit.

The halo case has a different metric hierarchy. When $a_0\gg r_\mathrm{c}$, Eqs.~\eqref{eq:metric_functions} imply that in the region $r\sim r_\mathrm{c}$,
%%%%%=====
\begin{align}
  \frac{\delta f_1}{f}
  &=-\frac{2M_\mathrm{h}}{a_0}
  \left[1+\mathcal{O}\!\left(\frac{r}{a_0}\right)\right],
  \\
  \frac{\delta f_2}{f}
  &=\mathcal{O}\!\left(\frac{M_\mathrm{h}r}{a_0^2}\right).
\end{align}
%%%%%
Thus the near-$r_c$ metric takes the form
%%%%%=====
\begin{align}
  ds^2
  \simeq
  -f\left(1-\frac{2M_\mathrm{h}}{a_0}\right)dt^2
  +\frac{dr^2}{f}
  +r^2d\Omega^2 .
\end{align}
%%%%%
The leading halo effect is therefore a constant time redshift, while the spatial metric differs from Schwarzschild only at relative order $r_\mathrm{c}/a_0$ compared with this redshift. The Hamiltonian potential retains precisely this leading lapse perturbation, $\delta V_\mathrm{h}\simeq -M_\mathrm{h}/a_0$, so it captures the leading large-$a_0$ halo effect.

The same diagnostic can be applied to the binary companion case. For a companion sitting at an orbital radius $R_*$, when $R_*\gg r_\mathrm{c}$, most of the scalar eigenfunctions lie inside the companion orbit and Eq.~\eqref{eq:delta_V_companions} gives, for $r\sim r_\mathrm{c}$,
%%%%%=====
\begin{align}
  \delta V_*
  =-\frac{qM_\mathrm{BH}}{R_*}
  +\mathcal{O}\!\left(\frac{qM_\mathrm{BH}r^2}{R_*^3}\right).
\end{align}
%%%%%
The leading $l_*=0$ term is therefore a constant redshift. The monopolar metric perturbation of a point particle, reconstructed for example in the Zerilli gauge as in Ref.~\cite{Brito:2023pyl}, does not shift the local Schwarzschild mass inside the orbit. Thus there is no disk-like spatial rescaling at the same order, and the Hamiltonian formula should capture the leading large-$R_*$ shift,
%%%%%=====
\begin{align}\label{eq:shift_analytical}
  \left(\frac{\delta\omega}{\omega}\right)_{l_*=0}
  \simeq -\frac{qM_\mathrm{BH}}{R_*},
\end{align}
%%%%%
for both the real and imaginary parts of the frequency.

When $R_*\sim r_\mathrm{c}\gg M_\mathrm{BH}$, the monopole and higher tidal multipoles are no longer parametrically ordered over the cloud. The metric perturbations due to a companion then contain lapse and spatial components with comparable amplitude. However, as in the disk case, the spatial-metric part enters the Klein--Gordon operator through spatial derivatives of the eigenfunctions, so its contribution is suppressed relative to the leading lapse coupling by $\alpha^2/\bar n^2$. The Hamiltonian formulation is therefore still appropriate for the leading companion-induced eigenfrequency shifts in this regime.

\subsection{First-order shift due to a binary companion}\label{sec:companion}
%%%%%FFFFF
\begin{figure*}[htb]
  \includegraphics[width=0.9\linewidth]{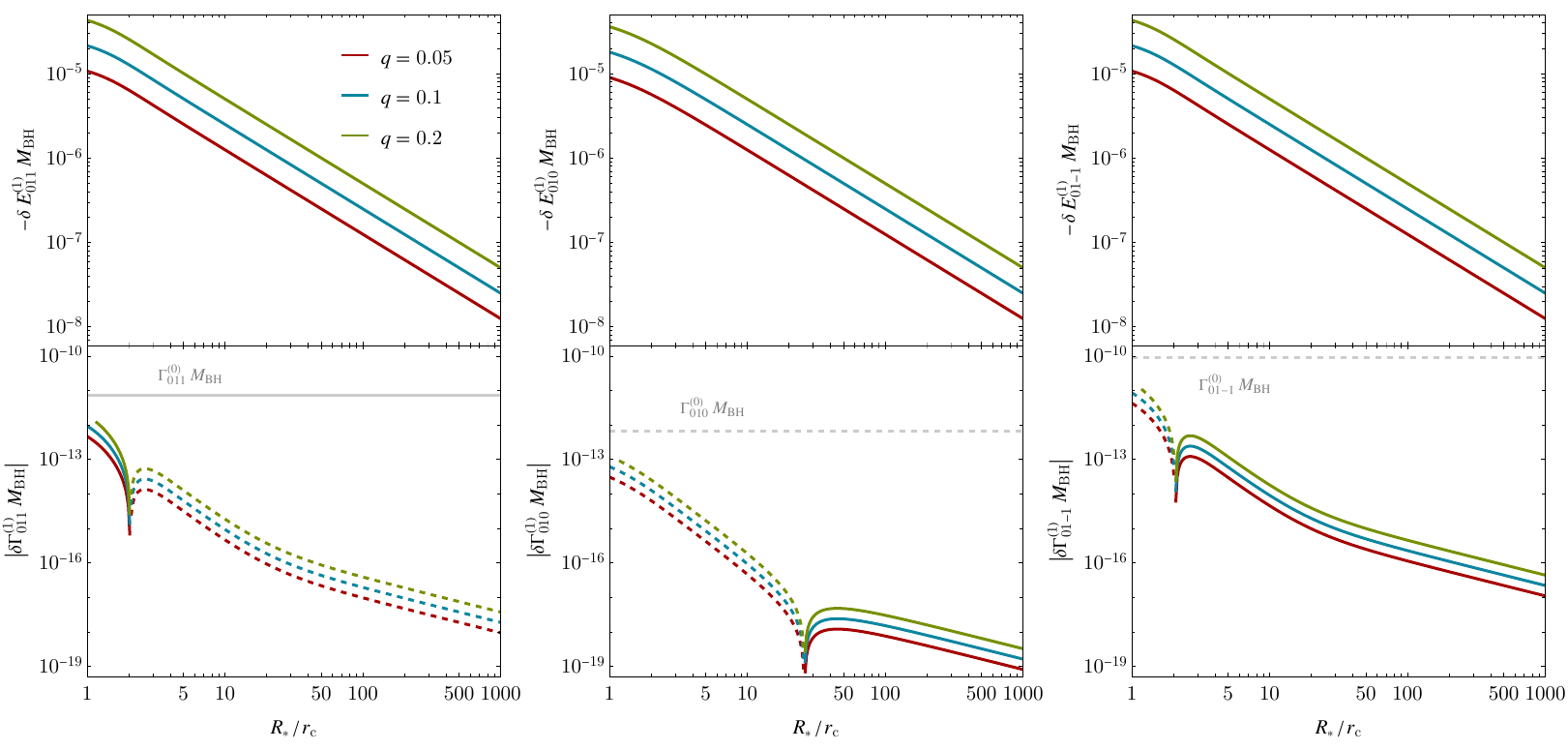}
  \caption{Corrections to the energy levels (top) and superradiant/decay rates (bottom) due to a binary companion, as functions of the binary separation $R_*$. The results are obtained using the Hamiltonian formulation of relativistic perturbation theory, Eq.~\eqref{eq:frequency_shift}, in the Kerr background. The left, middle, and right panels correspond to shifts for the $\{0,1,1\}$, $\{0,1,0\}$, and $\{0,1,-1\}$ modes, respectively. The red, blue, and green curves represent $q = 0.05$, $0.1$, and $0.2$, respectively. The other parameters are set to $a_* = 0.99$ and $\alpha = 0.1$. For the bottom panels, solid curves denote $\delta\Gamma^{(1)}_{nlm}>0$, dashed curves indicate $\delta\Gamma^{(1)}_{nlm}<0$, while the grey horizontal lines mark the (absolute) value of the zeroth-order superradiant/decay rates $\Gamma^{(0)}_{nlm}$.}
  \label{fig:companion_l=1}
\end{figure*}
%%%%%

Having assessed the validity of the Hamiltonian framework for the binary companion case, we can now adopt it to compute first-order shifts. As discussed in Sec.~\ref{subsec:binarycompanion}, we take the tidal potential to be given by~\eqref{eq:delta_V_companions} for binary separations $R_*\gtrsim r_c$, where the semi-Newtonian approximation should be reasonably accurate and the tidal potential can be used even for nearly equal-mass binaries. We also only consider circular, equatorial orbits, where $R_*$ is constant, $\theta_*=\pi/2$ and $\varphi_*=\Omega_* t$. Here we consider that the dimensionless BH spin is $a_*=0.99$ since we are interested in comparing our results with previous work~\cite{Tong:2022bbl} where, using a non-relativistic treatment, it was shown that a binary companion can turn a superradiantly unstable system stable, below a critical binary separation. For concreteness, we also focus on $\alpha=0.1$ but qualitatively similar results can be obtained for other values of $a_*$ and $\alpha$.

Our results are summarized in Fig.~\ref{fig:companion_l=1}. The top panels show the first-order shifts to the energy levels as functions of the binary separation $R_*$ for different modes and values of the binary mass ratio $q$. On the other hand, the bottom panels show the first-order shifts to the superradiant/decay rates~--~for this choice of parameters $\Gamma_{011}^{(0)}>0$ while $\Gamma_{010}^{(0)}<0$ and $\Gamma_{01-1}^{(0)}<0$~--~also as functions of the binary separation $R_*$ and for the same modes and mass ratios as the top panels.
For the shifts in the energy levels, the results of the non-relativistic perturbation method almost overlap with those of the relativistic perturbation method, and are therefore not shown in this figure. On the other hand, we remind that the shifts to the superradiant/decay rates are not captured by a non-relativistic approach.

We find that the energy shifts have the expected behavior; as the separation decreases or $q$ increases, the value of $|\delta E_{nlm}^{(1)}|$ increases, with the expected $\sim 1/R_*$ at large $R_*$ (cf. Eq.~\eqref{eq:shift_analytical}). On the other hand, the superradiant/decay-rate shifts have a more interesting behavior. When the binary separation $R_*$ is much larger than $r_c$, the sign of the correction is opposite to that of the unperturbed superradiant or decay rate. As the binary separation decreases, the absolute magnitude of the correction increases. Near $R_*\sim r_\mathrm{c}$ for the $l=|m|$ modes, or at separations of order $\mathcal{O}(10r_\mathrm{c})$ for the $l\neq|m|$ modes, the absolute magnitude of the correction rapidly decreases and the correction changes sign. Subsequently, the absolute magnitude of the correction increases again as the binary separation continues to decrease. 

It is worth noting that the magnitudes of the corrections $\delta \Gamma_{nlm}^{(1)}$ remain smaller than those of the unperturbed superradiant/decay rates throughout. In other words, to leading order in the perturbing potential, the presence of the companion does not turn superradiant (decaying) modes into decaying (superradiant) ones. This conclusion is not in contradiction with Ref.~\cite{Tong:2022bbl}, given that the change of sign of the superradiant rates found there is related to a mixing between superradiant and non-superradiant states which only occurs at second order in perturbation theory. Therefore, checking the results of Ref.~\cite{Tong:2022bbl} using relativistic perturbation theory requires computing second-order shifts to the eigenfrequencies. In the next section we extend the covariant relativistic perturbation theory method up to second order, using the galactic halo environment case as an example to check the validity of the method as well as discuss the limitations of using non-relativistic perturbation theory in this context.

\section{Second-order corrections and the completeness assumption}
\label{sec:second_order}

The results presented so far are based on first-order perturbation theory. For environments which preserve axial symmetry, such as the galactic halo and the disk environments, higher-order corrections are not expected to modify the conclusions obtained at leading order. However, binary companions break axisymmetry and can therefore induce couplings between superradiant and stable modes~\cite{Tong:2022bbl}. Since this mode mixing arises at second order, beyond the first-order self-energy correction, it must be investigated separately. In this section we summarize the formalism to compute second-order eigenfrequency shifts, following Ref.~\cite{Lestingi:2026peq}, and benchmark it with the non-perturbative numerical solutions obtained for the galactic halo environment. We should note that, since the halo is spherically symmetric, mode mixing can only occur between overtones with the same angular numbers, and therefore no mixing between different $m$-modes arises in this setup.

\subsection{Formalism}

According to Ref.~\cite{Lestingi:2026peq}, the second-order shifts to the eigenfrequencies in the covariant formalism can be written as:
%%%%%=====
\begin{align}\label{eq:2nd_order_shift}
    \begin{split}
      \delta \omega_{nlm}^{(2)} = \frac{i (\delta \mathcal{O}^{(2)})_{{nlm},{nlm}}}{\langle\langle \Phi_{nlm}^{(0)}, \Phi_{nlm}^{(0)} \rangle\rangle} - \delta \omega_{nlm}^{(1)} \frac{\langle\langle \Phi_{nlm}^{(0)}, \delta \Phi_{nlm}^{(1)} \rangle\rangle}{\langle\langle \Phi_{nlm}^{(0)}, \Phi_{nlm}^{(0)} \rangle\rangle} &
      \\
      + \frac{i}{\langle\langle \Phi_{nlm}^{(0)}, \Phi_{nlm}^{(0)} \rangle\rangle} \int_{\mathcal{C}} \left(\mathcal{J}\Phi_{nlm}^{(0)}\right) \left( \delta \mathcal{O}^{(1)} \delta \Phi_{nlm}^{(1)} \right) t^\mu d\Sigma_\mu, &
    \end{split}
\end{align}
%%%%%
where $t^\mu$ is the Kerr or Schwarzschild spacetime timelike Killing vector field and $\mathcal{C}$ denotes the complex contour used in the bilinear form (see Sec.~\ref{sec:perturbation_theory}). 
For the galactic halo environment, the second-order operator is explicitly given by
%%%%%=====
\begin{align}\label{eq:halo_second_order_operator}
    \begin{split}
      f(r)\delta\mathcal{O}^{(2)}
      =\frac{4 M_\mathrm{h}^2 (2 a_0+3 r-2 M_\mathrm{BH})}{3 (a_0+r)^3}\partial_t^2 .
    \end{split}
\end{align}
%%%%%

The main difficulty in computing $\delta \omega_{nlm}^{(2)}$ is the fact that Eq.~\eqref{eq:2nd_order_shift} explicitly depends on the first-order perturbed field $\delta \Phi^{(1)}$. This quantity can be formally defined by expanding the full solution. Recalling that we defined the small parameter of the expansion as $\epsilon_{\rm h} \equiv M_{\rm h}/a_0$, we can write the full solution separating the harmonic time-dependence from the time-independent part
\begin{equation}
 \Phi_{\rm full} =e^{-i \omega(\epsilon_{\rm h}) t}\chi(r, \theta, \varphi, \epsilon_{\rm h}) \, ,
\end{equation}
so that the first-order solution reads:
%%%%%=====
\begin{align}\label{eq:first_order_mode_time_dependence}
  \delta\Phi_{{nlm}}^{(1)}(t,\mathbf{r}) = e^{-i\omega_{{nlm}}^{(0)}t}\,\chi_{{nlm}}^{(1)}(\mathbf{r}) - i\delta\omega_{{nlm}}^{(1)}t\,\Phi_{{nlm}}^{(0)}(t,\mathbf{r}).
\end{align}
%%%%%
Correspondingly, we define $\chi_{nlm}^{(0)}\equiv e^{i\omega_{nlm}^{(0)}t}\Phi_{nlm}^{(0)}$.
As evident from this equation, the first-order solution has a component which grows linearly with time $t$. Nevertheless, once this expression is inserted into Eq.~\eqref{eq:2nd_order_shift}, this contribution vanishes exactly~\cite{Lestingi:2026peq}, leaving a well-defined second-order shift. In addition, since $\delta\mathcal{O}^{(1)}$ in general contains second-order time derivatives [see Eq.~\eqref{eq:Operator_1storder_halo}], it is useful to decompose it as
%%%%%=====
\begin{align}
  \delta\mathcal{O}^{(1)}
  =\delta\mathcal{O}_{\rm sp}^{(1)}+\beta(r)\partial_t^2,
\end{align}
%%%%%
where $\delta\mathcal{O}_{\rm sp}^{(1)}$ denote the terms containing only spatial derivatives and $\beta(r)=2M_\mathrm{h}/[(a_0+r)f(r)]$ for the galactic halo case. In the frequency-domain we can therefore write $\delta\mathcal{O}_{\rm freq}^{(1)}(\omega)\equiv\delta\mathcal{O}_{\rm sp}^{(1)}-\beta(r)\omega^2$ obtained via the substitution $\partial_t^2\rightarrow-\omega^2$.

To further simplify Eq.~\eqref{eq:2nd_order_shift}, we introduce the following definitions:
%%%%%=====
\begin{subequations}
\begin{align}
  \mathcal{F}_{nlm}
  &\equiv
  f(r)\delta\mathcal{O}_{\rm freq}^{(1)}(\omega_{nlm}^{(0)}),
  \\
  \mathsf{C}_{nlm}
  &\equiv
  \frac{\langle\langle\Phi_{nlm}^{(0)},f(r)\delta\mathcal{O}^{(2)}
  \Phi_{nlm}^{(0)}\rangle\rangle}
  {2\omega_{nlm}^{(0)}
  \langle\langle\Phi_{nlm}^{(0)},\Phi_{nlm}^{(0)}\rangle\rangle}
  -\frac{(\delta\omega_{nlm}^{(1)})^2}{2\omega_{nlm}^{(0)}}
  \nonumber\\
  &\hspace{1cm}
  -\delta\omega_{nlm}^{(1)}
  \frac{\langle\langle\Phi_{nlm}^{(0)},f(r)\beta(r)\Phi_{nlm}^{(0)}\rangle\rangle}
  {\langle\langle\Phi_{nlm}^{(0)},\Phi_{nlm}^{(0)}\rangle\rangle},
  \\
  \mathcal{K}_{nlm}[\chi]
  &\equiv
  \frac{\langle\langle\Phi_{nlm}^{(0)},e^{-i\omega_{nlm}^{(0)}t}
  \mathcal{F}_{nlm}\chi\rangle\rangle}
  {2\omega_{nlm}^{(0)}
  \langle\langle\Phi_{nlm}^{(0)},\Phi_{nlm}^{(0)}\rangle\rangle}
  \nonumber\\
  &\hspace{1cm}
  -\delta\omega_{nlm}^{(1)}
  \frac{\langle\langle\Phi_{nlm}^{(0)},e^{-i\omega_{nlm}^{(0)}t}
  \chi\rangle\rangle}
  {\langle\langle\Phi_{nlm}^{(0)},\Phi_{nlm}^{(0)}\rangle\rangle} .
\end{align}
\end{subequations}
%%%%%
The second-order shift can then be written in the compact form:
%%%%%=====
\begin{align}\label{eq:second_order_without_completeness}
  \delta\omega_{nlm}^{(2)}
  =
  \mathsf{C}_{nlm}
  +\mathcal{K}_{nlm}[\chi_{nlm}^{(1)}] \, ,
\end{align}
%%%%%
where we recall that $\chi_{nlm}^{(1)}$ is the time-independent first-order perturbation, i.e. the correction to the radial wavefunction.

In principle, $\chi_{nlm}^{(1)}$ can be computed by expanding the perturbative equations to first-order and solving the resulting differential equation [Eq.~\eqref{eq:KG_first_order_app} in App.~\ref{appendix:linear_system_derivation}]. We will not do this here, since our main goal is to check the accuracy of the formalism to compute frequency shifts. Since in the galactic halo environment case we can compute the full solution $\chi_{nlm}$ by directly integrating Eq.~\eqref{eq:radial_eq_galaxy} with appropriate boundary conditions, $\chi_{nlm}^{(1)}$ can be obtained by subtracting the zeroth-order solution $\chi_{nlm}^{(0)}$ from $\chi_{nlm}$:
%%%%%=====
\begin{align}
    \chi_{nlm}^{(1)} \simeq \chi_{nlm}-\chi_{nlm}^{(0)}.
\end{align}
%%%%%
In the limit $\epsilon_{\rm h}\to 0$ this provides a controlled determination of $\chi_{nlm}^{(1)}$. 

We aim to contrast this procedure to what would be done if we assumed a Hermitian system. In this case, as in ordinary quantum mechanics, one could instead assume that the spatial quasibound states form a complete basis of a Hilbert space and reconstruct $\chi_{nlm}^{(1)}$ from a linear combination of zeroth-order wavefunctions. Under this assumption, one can write
%%%%%=====
\begin{align}\label{eq:chi_spatial_expansion}
  \chi_{nlm}^{(1)}
  =\sum_{\mathbf{p}\in\mathcal{B}_{nlm}}
  c_{\mathbf{p}}\chi_{\mathbf{p}}^{(0)},
\end{align}
%%%%%
where $\mathcal{B}_{nlm}$ excludes the subspace spanned by the eigenstates degenerate with the reference $|nlm\rangle$ mode.
Defining the projector
%%%%%=====
\begin{align}\label{eq:projetor}
  \mathcal{P}_{\mathbf{p}}[\chi]\equiv
  \langle\langle
  \Phi_{\mathbf{p}}^{(0)},e^{-i\omega_{\mathbf{p}}^{(0)}t}\chi
  \rangle\rangle ,
\end{align}
%%%%%
inserting the expansion~\eqref{eq:chi_spatial_expansion} into the first-order Klein-Gordon equation and projecting onto the basis yields a linear system for the coefficients $c_{\mathbf{p}}$ (see Appendix~\ref{appendix:linear_system_derivation} for a detailed derivation). Denoting $\mathcal{N}_{\mathbf{p}}\equiv\langle\langle\Phi_{\mathbf{p}}^{(0)},\Phi_{\mathbf{p}}^{(0)}\rangle\rangle$, the orthogonality of the unperturbed modes in the Schwarzschild background reads 
%%%%%=====
\begin{align}\label{eq:ortho_main}
  \mathcal{P}_{\mathbf{p}}[\chi_{\mathbf{q}}^{(0)}]
  =\mathcal{N}_{\mathbf{p}}\,\delta_{\mathbf{p}\mathbf{q}},
\end{align}
%%%%%
which follows from the orthogonality of the bilinear form \cite{Cannizzaro:2023jle}. This allows us to solve for the coefficients as
%%%%%=====
\begin{align}\label{eq:c_decoupled_main}
  c_{\mathbf{p}}=-\,\frac{\mathcal{P}_{\mathbf{p}}\!\left[\mathcal{F}_{nlm}\chi_{nlm}^{(0)}\right]}
  {\left[(\omega_{\mathbf{p}}^{(0)})^2-(\omega_{nlm}^{(0)})^2\right]\mathcal{N}_{\mathbf{p}}},
  \qquad \mathbf{p}\in\mathcal{B}_{nlm} .
\end{align}
%%%%%
Substituting Eq.~\eqref{eq:c_decoupled_main} into Eq.~\eqref{eq:second_order_without_completeness} and using the orthogonality~\eqref{eq:ortho_main} to eliminate the cross-term inside $\mathcal{K}_{nlm}$, the second-order frequency shift under the completeness assumption reads
%%%%%=====
\begin{align}\label{eq:second_order_spectral}
  \begin{split}
    \delta\omega_{nlm}^{(2)}\big|_{\rm comp}
    &=\mathsf{C}_{nlm}
    \\
    &\hspace{-1cm}-\frac{1}{2\omega_{nlm}^{(0)}\mathcal{N}_{nlm}}
    \sum_{\mathbf{p}\in\mathcal{B}_{nlm}}
    \frac{\mathsf{V}_{nlm,\mathbf{p}}\,\mathsf{V}_{\mathbf{p},nlm}}
    {\left[(\omega_{\mathbf{p}}^{(0)})^2-(\omega_{nlm}^{(0)})^2\right]\mathcal{N}_{\mathbf{p}}},
  \end{split}
\end{align}
%%%%%
where $\mathsf{V}_{\mathbf{a}\mathbf{b}}\equiv\mathcal{P}_{\mathbf{a}}\!\left[\mathcal{F}_{nlm}\chi_{\mathbf{b}}^{(0)}\right]$. 

Finally, following Ref.~\cite{Tong:2022bbl}, one may also try to estimate the second-order frequency shift through the standard non-relativistic spectral-sum formula. In this approach, the system is treated as effectively Hermitian, analogously to ordinary quantum mechanics, and the second-order correction is reconstructed assuming completeness of the zeroth-order spectrum. The dissipative nature of the relativistic problem is then incorporated only phenomenologically, by adding the imaginary parts of the frequencies {\it a posteriori}. For brevity, let $\mathbf{n}\equiv(nlm)$ and define
%%%%%=====
\begin{align}
  U_{\mathbf{a}\mathbf{b}}^{(i)}
  \equiv
  \langle\psi_{\mathbf{a}}^{(0)}|
  \mu_\mathrm{s}\delta V^{(i)}
  |\psi_{\mathbf{b}}^{(0)}\rangle .
\end{align}
%%%%%
The second-order non-relativistic expression is then
%%%%%=====
\begin{align}\label{eq:second_order_qm}
  \delta\omega_{\mathbf{n}}^{(2)}\big|_{\rm QM}
  =
  U_{\mathbf{n}\mathbf{n}}^{(2)}
  +\sum_{\mathbf{k}\neq\mathbf{n}}
  \frac{\big|U_{\mathbf{n}\mathbf{k}}^{(1)}\big|^2}
  {\omega_{\mathbf{n}}^{(0)} - \omega_{\mathbf{k}}^{(0)}} ,
\end{align}
%%%%%
with
%%%%%=====
\begin{align}
    \delta V^{(2)} = \frac{2(a_0+2M_\mathrm{BH})M_\mathrm{h}^2}{3(r+a_0)^3}\,,
\end{align}
%%%%%
for the galactic halo environment. Here, the sum is over the hydrogenic basis after resolving possible degeneracies in the usual way. 

\subsection{Results}

%%%%%FFFFF
\begin{figure*}[htb]
  \includegraphics[width=0.9\linewidth]{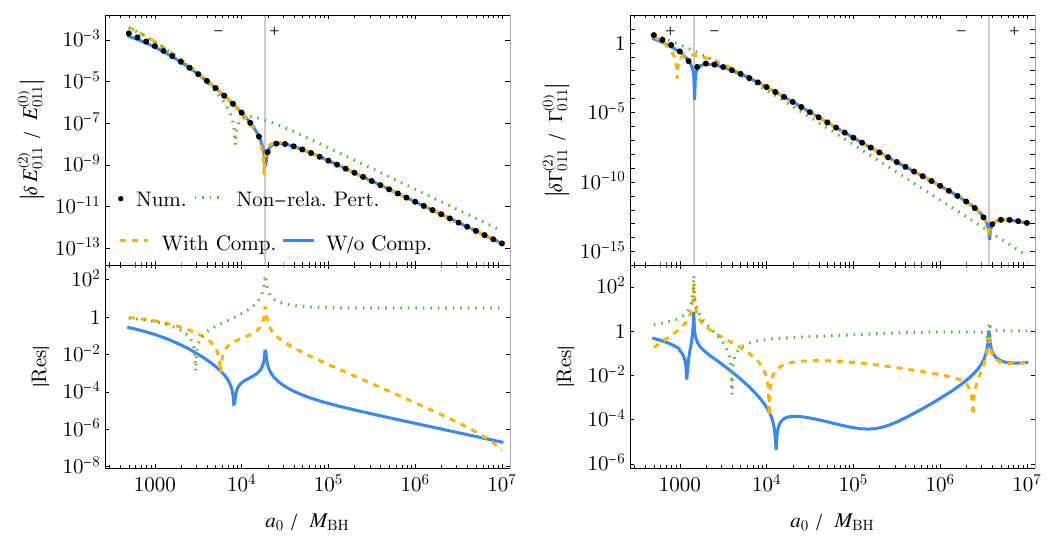}
  \caption{Second-order relative eigenfrequency shifts (top panels) and the corresponding absolute fractional residuals (bottom panels) as functions of $a_0/M_{\rm BH}$, for the $(n,l,m)=(0,1,1)$ mode. These results are obtained in a Schwarzschild background. Black points denote $(\omega_{nlm}-\omega_{nlm}^{(0)}-\delta\omega_{nlm}^{(1)})/\omega_{nlm}^{(0)}$, where $\omega_{nlm}$ is the numerical solution obtained by numerically integrating Eq.~\eqref{eq:radial_eq_galaxy}. The blue solid, orange dashed, and green dotted lines represent the results from the relativistic covariant formulation without assuming completeness [Eq.~\eqref{eq:second_order_without_completeness}], assuming completeness [Eq.~\eqref{eq:second_order_spectral}], and the non-relativistic perturbation method [Eq.~\eqref{eq:second_order_qm}], respectively. The left panels show the relative corrections and their corresponding absolute fractional residuals for the energy levels, while the right panels display those for the decay rates. In the top panels, the vertical line marks the zero-crossing for the numerical results. The other parameters are fixed at $M_\mathrm{h} = 10 M_\mathrm{BH}$ and $\alpha = 0.1$.}
  \label{fig:2nd_shift}
\end{figure*}
%%%%%

The results for second-order eigenfrequency shifts are shown in Fig.~\ref{fig:2nd_shift} where we compare the different approaches discussed above, against the non-perturbative numerical results. For concreteness we fix $\alpha=0.1$ and $M_{\rm h}=10M_{\rm BH}$. For the non-perturbative results, we define the shifts as $\omega_{nlm}-\omega_{nlm}^{(0)}-\delta\omega_{nlm}^{(1)}$, where $\omega_{nlm}$ is obtained by numerically integrating Eq.~\eqref{eq:radial_eq_galaxy}, while $\omega_{nlm}^{(0)}$ is the Schwarzschild quasibound state and $\delta\omega_{nlm}^{(1)}$ is computed using Eq.~\eqref{eq:frequency_shift_cov}. 
The second-order fractional residual is defined as
%%%%%=====
\begin{align}
    {\rm Res \equiv \frac{2nd\ order}{Num.} - 1}.
\end{align}
%%%%%

As expected, we find that the relativistic perturbative predictions agree very well with the non-perturbative numerical results when the completeness assumption is not imposed. On the other hand, when the completeness assumption is enforced, the residuals are systematically larger and the agreement deteriorates at low $a_0/M_{\rm BH}$, although the predictions remain reasonably accurate over a substantial range of parameters. By contrast, the non-relativistic spectral-sum formula exhibits sizably larger deviations for all values of $a_0/M_{\rm BH}$.

Importantly, the non-relativistic calculation fails to reproduce the numerical decay-rate shifts, yielding residuals of order unity or larger. This demonstrates that estimates based on the non-relativistic results should be taken with care. In contrast, the relativistic framework provides a much more accurate description of the spectrum. When completeness is not assumed, excellent agreement with the numerical results is obtained across all the cases considered.

The role of the completeness assumption within the relativistic framework is subtle. As shown in the figure, for quasibound states this assumption can provide a reasonable approximation when the perturbation predominantly overlaps with the asymptotic tail of the cloud, i.e. for large $a_0/M_{\rm BH}$. However, visible discrepancies emerge once the perturbation probes the bulk of the eigenfunction, where deviations from the asymptotic hydrogenic behavior are largest. In these cases, the completeness assumption fails to reproduce the correct decay-rate shifts, and even energy shifts, despite the underlying relativistic treatment.\footnote{It is worth pointing out that, although Eq.~\eqref{eq:second_order_spectral} does not reproduce well the eigenfrequency shifts in some parts of the parameter space, we find that the sum in that expression always converges. This is in contrast to what happens for quasinormal modes, where the analogue of Eq.~\eqref{eq:second_order_spectral} does not appear to ever converge~\cite{Lestingi:2026peq}.}

This behavior suggests that the accuracy of the completeness approximation depends on the region of the eigenfunction probed by the perturbation. While the asymptotic tail of the eigenfunctions appears to be well approximated by an expansion in zeroth-order quasibound states, the relativistic bulk at small values of $r$ should be more sensitive to the non-Hermitian character of the BH spectrum. Thus, we expect the breakdown of the completeness assumption to be associated with the inability of a complete quasibound-state expansion to fully capture the near-horizon physics responsible for dissipation.

To further test this interpretation, we repeated the analysis for $\alpha=0.3$. In this regime, the bulk of the eigenfunction is localized closer to the BH, so that for a fixed halo profile at $a_0=10^3 M_{\rm BH}$ the perturbation probes a larger fraction of the asymptotic tail and a smaller fraction of the relativistic bulk. Consistent with the above picture, the completeness-based calculation shows improved agreement with the numerical results. 
Specifically, for $a_0=10^3 M_{\rm BH}$, the absolute fractional residual of the completeness-based prediction for the energy shift is $1.70\times10^{-2}$ when $\alpha=0.3$ compared to $0.61$ when $\alpha=0.1$. For the decay-rate shift, the corresponding residual is $3.77\times10^{-3}$ for $\alpha=0.3$ compared to $1.26$ for $\alpha=0.1$.
This trend supports the interpretation that the observed discrepancies are tied to the overlap between the perturbation and the relativistic region of the cloud.

Overall, these findings indicate that completeness can remain a useful approximation in specific regimes, but its domain of validity should be studied with care. Accurately reproducing the full relativistic spectrum across the entire parameter space, including decay and superradiant rates, requires a fully relativistic treatment that does not rely on completeness assumptions.
Notably, these results indicate that previous estimates of environmental effects on superradiance based on non-relativistic treatments~\cite{Tong:2022bbl, Li:2026gup} should be revisited within a relativistic calculation, in order to assess their robustness. In particular, such approaches miss the first-order correction to the imaginary part of the spectrum and can fail to correctly reproduce the second-order shifts, due to the completeness assumption and the {\it a posteriori} inclusion of dissipative effects.  

\section{Summary and discussion} \label{sec:summary}
In this work, we developed and applied a relativistic perturbation theory to study how different environments modify the quasibound-state spectrum of massive scalar fields around BHs. As examples, we considered galactic halos, accretion disks, and binary companions, and used halos and disks to benchmark relativistic perturbation theory against non-perturbative numerical solutions to the eigenfrequencies. At first order, we showed that a fully relativistic treatment captures corrections to both the real and imaginary parts of the spectrum, while non-relativistic approaches fail to capture the decay-rate shift. We then extended the perturbative analysis to second order, where mode mixing enters, and used the galactic halo case to test the completeness assumption. We found that completeness-based approaches fail to fully reproduce the dissipative spectrum in parts of the parameter space, in agreement with the fact that quasibound states do not form a complete basis of the relativistic problem. Therefore, following Ref.~\cite{Lestingi:2026peq}, we formulated a consistent relativistic treatment capturing both dissipative effects and mode mixing without relying on completeness assumptions.

Our results show that previous estimates, based on non-relativistic perturbation theory, for the onset and termination of superradiance in the presence of binaries or (non-axisymmetric) accretion disks~\cite{Tong:2022bbl, Li:2026gup} should be revisited within our fully relativistic framework.
However, performing this analysis for companions requires determining the first-order correction to the eigenfunctions without relying on completeness. This can in principle be done using a small mass-ratio approximation, following Refs.~\cite{Brito:2023pyl,Dyson:2025dlj,Xu:2026aic,Li:2025ffh}. We leave these developments for future work.

\section*{Acknowledgments}
We thank Vitor Cardoso for useful comments and feedback on the manuscript. 
We thank Jacopo Lestingi for his valuable help and insightful discussions on the second-order calculations.
Y.G. is grateful to Vitor Cardoso and R.B. for their warm hospitality during his stay at CENTRA/IST. He acknowledges financial support from the National Natural Science Foundation of China (Grants Nos. 124B2098, 12447105, 12075136) and from the Natural Science Foundation of Shandong Province (Grant No. ZR2020MA094). 
R.B. and Q.X. acknowledge financial support provided by FCT – Fundação para a Ciência e a Tecnologia, I.P., through the ERC-Portugal program Project ``GravNewFields''. 
Q.X. also acknowledges support from FCT through grant \href{https://doi.org/10.54499/2025.01546.BD}{2025.01546.BD}.
E.C. acknowledges financial support provided under the European Union’s H2020 ERC Advanced Grant “Black holes: gravitational engines of discovery” grant agreement no. Gravitas–101052587, and from the Villum Investigator program supported by the VILLUM Foundation (grant no. VIL37766) and the DNRF Chair program (grant no. DNRF162) by the Danish National Research Foundation.
We also thank the Fundação para a Ciência e Tecnologia (FCT), Portugal, for the financial support to the Center for Astrophysics and Gravitation (CENTRA/IST/ULisboa) through grant No.~\href{https://doi.org/10.54499/UID/PRR/00099/2025}{UID/PRR/00099/2025} and grant No.~\href{https://doi.org/10.54499/UID/00099/2025}{UID/00099/2025}.

% \appendix

\appendix

%\clearpage
%%%%%%%%%%%%%%%%%%%%%%%%%%%%%%%%%%%%
\section{The projection method for the Schwarzschild BH-disk model}
\label{appendix:projection_method}
%%%%%%%%%%%%%%%%%%%%%%%%%%%%%%%%%%%%
%%%%%
In this appendix we give more details about the approximation we employ in Sec.~\ref{sec:Accretion} and define the coefficients and functions appearing in Eq.~\eqref{eq:radial_eq}.

We first perform a Taylor expansion of the functions appearing in the metric~\eqref{eq:metric_disk} in terms of $x \equiv \cos\theta$:
\begin{equation}\label{eq:Taylor_expand}
\begin{aligned}
\nu_\mathrm{disk} &= \epsilon_{\rm d} \mathcal{V}_{j}(r) |x^j|,
\qquad
\lambda_\mathrm{ext} &= \epsilon_{\rm d} \mathcal{L}_{j}(r) |x^j|\,,
\end{aligned}
\end{equation}
where $\mathcal{V}_{j}(r)$ and $\mathcal{L}_{j}(r)$ are radial functions and a summation over $j$ is implicitly assumed. In the limit $\epsilon_{\rm d} \ll 1$, the metric~\eqref{eq:metric_disk} reduces to the approximate form
%%%%%=====
\begin{subequations}
  \begin{align}
    g_{tt} & =-\left(1-\frac{2M_\mathrm{BH}}{r}\right)\left(1+\epsilon_{\rm d} A_{j}(r)|\cos^{j}\theta|\right),\\
    g_{rr} & =\left(1-\frac{2M_\mathrm{BH}}{r}\right)^{-1}\left(1+\epsilon_{\rm d} B_{j}(r)|\cos^{j}\theta|\right),\\
    g_{\theta\theta} & =r^{2}\left(1+\epsilon_{\rm d} C_{j}(r) |\cos^{j}\theta| \right),\\
    g_{\varphi\varphi} & =r^{2}\sin^{2}\theta\left(1+\epsilon_{\rm d} D_{j}(r) |\cos^{j}\theta| \right),
  \end{align}
\end{subequations}
%%%%%
where 
%%%%%=====
\begin{subequations}
  \begin{align}
    A_j(r) & = -D_j(r) = 2\mathcal{V}_{j}(r),\\
    B_j(r) & = C_j(r) = 2\mathcal{L}_{j}(r) - 2\mathcal{V}_{j}(r).
  \end{align}
\end{subequations}
%%%%%
By substituting this metric into the Klein-Gordon equation and following the projection procedure shown in Ref.~\cite{Chen:2022ynz}, we find Eq.~\eqref{eq:radial_eq} with
%%%%%=====
\begin{align}
  U_{lm}(r) & =2 \mathcal{V}_j a_{lm}^{j}  -\left(2 \mathcal{V}_j - 2 \mathcal{L}_j\right)c_{lm}^{j},
\end{align}
%%%%%
and
%%%%%=====
\begin{subequations}
  \begin{align}
    a_{lm}^j &= \frac{2{m}^2}{\mathsf{n}_{lm}}\int_0^1\frac{x^j\left(P_l^{m}\right)^2}{1-x^2}dx,
    \\
    b_{lm}^j &= \frac{2}{\mathsf{n}_{lm}}\int_0^1x^j\left(P_l^{m}\right)^2dx, \\
    c_{lm}^j &= \frac{2}{\mathsf{n}_{lm}}\int_0^1x^jP_l^{m}\left[\left(1-x^2\right)\partial_x^2-2x\partial_x\right]P_l^{m}dx\,. 
    % \\
    % d_{lm}^j &= \frac{2}{\mathsf{n}_{lm}}\int_0^1P_l^{m}\left(1-x^2\right)\left(\partial_xx^j\right)\left(\partial_xP_l^{m}\right)dx.
  \end{align}
\end{subequations}
%%%%%
Here, $P_l^{m}(x)$ is the associated Legendre function, and $\mathsf{n}_{lm}$ represents its normalization constant, defined as
%%%%%=====
\begin{align}
  \int_{-1}^{1}dx P_l^m(x) P_{l'}^m(x) = \frac{2(l+m)!}{(2l+1)(l-m)!} \delta_{ll'} \equiv \mathsf{n}_{lm} \delta_{ll'}.
\end{align}
%%%%%
\section{Counter-term regularization in Kerr}
\label{appendix:counterterm}
%%%%%
In this appendix we present the derivation of the regularized form of the bilinear product in a Kerr background [see Eq.~\eqref{eq:regular_bilinearform}]. 

Near the horizon, the radial part of the quasibound solution admits the asymptotic form
%%%%%=====
\begin{align}
    \lim_{r_*\rightarrow-\infty} R = C e^{-ik_\mathrm{H}r_*},
\end{align}
%%%%%
where $C$ is a constant and $k_\mathrm{H} = \omega^{(0)} - m \Omega_\mathrm{H}$. Here, $r_*$ is the tortoise coordinate, defined through Eq.~\eqref{eq:tortoise}.
%%%%%=====
%\begin{align}
%    \frac{dr}{dr_*} = \frac{\Delta}{r^2+a^2}.
%\end{align}
%%%%%
Changing variables from $r$ to $r_*$, the radial integral can be rewritten as
%%%%%=====
\begin{align}
    \int\limits_{\mathcal{C}} dr\,\frac{K}{\Delta} R_1 R_2 = \int\limits_{\mathcal{C}} dr_*\,\frac{K}{r^2+a^2} R_1 R_2\,.
\end{align}
%%%%%
This last integral can be decomposed into
%%%%%=====
\begin{align}
    \begin{split}
        &\int\limits_{\mathcal{C}} dr_*\frac{K}{r^2+a^2} R_1 R_2 =
        \\
        &\lim_{\bar{r}_* \rightarrow -\infty} \left(\frac{K(\bar{r})}{\bar{r}^2+a^2}\int\limits_{\mathcal{C}_+} dr_*R_1 R_2 + \int_{\bar{r}_*}^\infty dr_*\,\frac{K}{r^2+a^2} R_1 R_2\right),
    \end{split}
\end{align}
%%%%%
where the deformed near-horizon contour is parameterized as $\mathcal{C}_+:r_*=\bar{r}_*+\rho e^{i\beta}$, $\beta$ is an angle in the complex plane that satisfies Eq.~\eqref{eq:arg_contour} and $\bar{r}\equiv r(\bar{r}_*)$. The first term can be analytically continued and evaluated explicitly in the limit $\bar{r}_*\to -\infty$:
%%%%%=====
\begin{align}
    \begin{split}
        &\frac{K(\bar{r})}{\bar{r}^2+a^2}\int\limits_{\mathcal{C}_+} dr_* R_1 R_2 
        \\
        =& \frac{K(\bar{r})}{\bar{r}^2+a^2}C_1 C_2 e^{i\beta}\int^{0}_\infty d\rho e^{-i[\omega_{1}^{(0)}+\omega_{2}^{(0)} - (m_1+m_2) \Omega_\mathrm{H}](\bar{r}_*+\rho e^{i\beta})}
        \\
        =& \frac{K(\bar{r})}{\bar{r}^2+a^2} \frac{iR_1(\bar{r}) R_2(\bar{r})}{\omega_{1}^{(0)}+\omega_{2}^{(0)} - (m_1+m_2) \Omega_\mathrm{H}}.
    \end{split}
\end{align}
%%%%%
Substituting this result back into Eq.~\eqref{eq:bilinear_kerr}, we finally obtain Eq.~\eqref{eq:regular_bilinearform} which, in the  Schwarzschild limit and after performing the integrations over $\theta$, reduces to the result in Ref.~\cite{Cannizzaro:2023jle}:
%%%%%=====
\begin{align}
  \begin{split}
    & \langle\langle\Phi_1,\Phi_2\rangle\rangle = -2\pi i \delta_{l_1 l_2}\delta_{m_1 m_2} (\omega_{1}^{(0)}+\omega_{2}^{(0)}) e^{i(\omega_{1}^{(0)}-\omega_{2}^{(0)})t}\times
    \\
    & \lim_{\bar{r}\rightarrow 2M_\mathrm{BH}}\Bigg[ \int_{\bar{r}}^{\infty} \frac{r^2 dr}{f(r)} R_1(r)R_2(r) + \frac{i \bar{r}^2}{\omega_1+\omega_2}  R_1(\bar{r})R_2(\bar{r}) \Bigg],
  \end{split}
\end{align}
%%%%%
where $f(r)=1-2M_\mathrm{BH}/r$.

\section{First-order shifts for the accretion disk model}
\label{sec:disk_shifts}
%%%%%FFFFF
\begin{figure*}
  \includegraphics[width=0.9\linewidth]{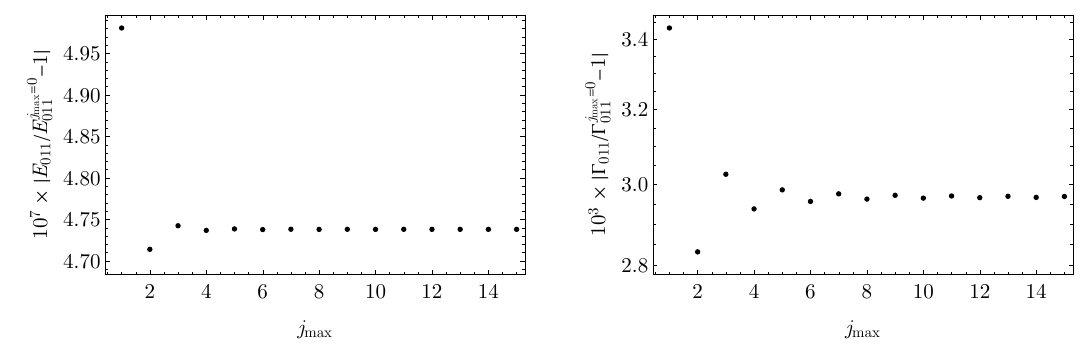}
  \caption{Convergence tests for $| E_{011}/E_{011}^{j_\mathrm{max}=0} - 1|$ and $|\Gamma_{011}/\Gamma_{011}^{j_\mathrm{max}=0}-1|$ as functions of $j_\mathrm{max}$ for numerically computed eigenfrequencies in the BH-disk model. The parameters are set to $b = 10 M_\mathrm{BH}$, $\epsilon_{\rm d} = 0.01$, and $\alpha = 0.1$.}
  \label{fig:vs_jmax}
\end{figure*}
%%%%%

%%%%%FFFFF
\begin{figure*}[htb]
  \includegraphics[width=0.9\linewidth]{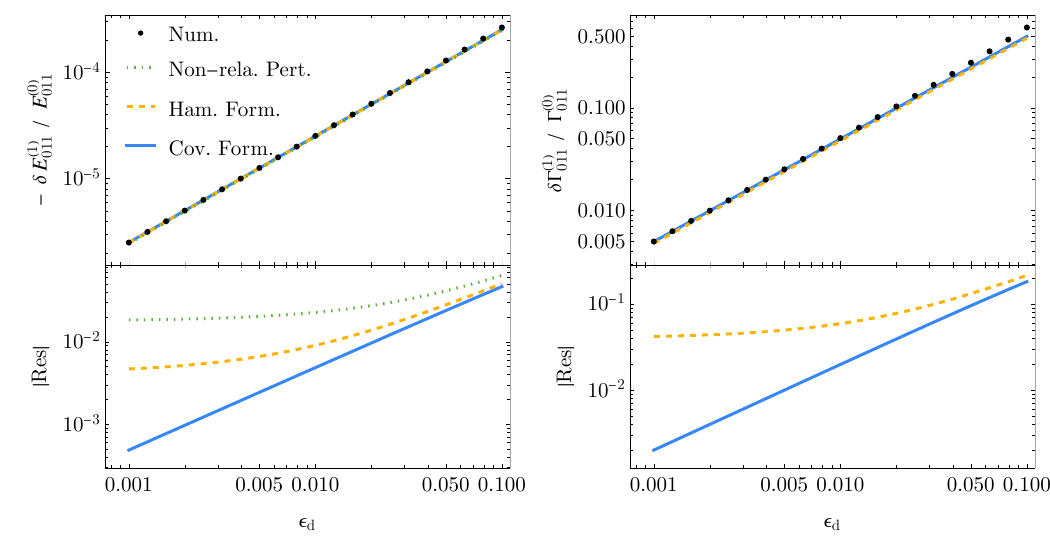}
  \caption{Same as Fig.~\ref{fig:diff_b}, but now showing results as a function of $\epsilon_{\rm d}$. The other parameters are fixed at $b = 10 M_\mathrm{BH}$ and $\alpha = 0.1$.}
  \label{fig:diff_epsilon}
\end{figure*}
%%%%%
%%%%%FFFFF
\begin{figure*}[htb]
  \includegraphics[width=0.9\linewidth]{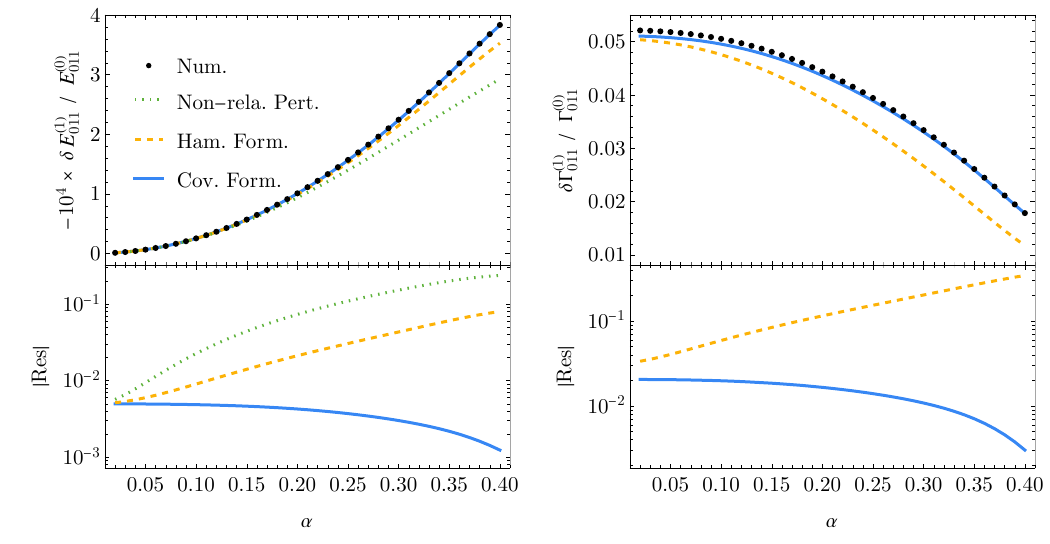}
  \caption{
  Same as Fig.~\ref{fig:diff_b}, but now showing results as a function of $\alpha$. The other parameters are fixed at $\epsilon_{\rm d} = 0.01$ and $b = 10M_\mathrm{BH}$.}
  \label{fig:diff_mu_disk}
\end{figure*}
%%%%%

In this appendix we present more details, concerning the first-order eigenfrequency shifts for the accretion disk case, regarding convergence and dependence on additional parameters of the model not discussed in the main text. 

The convergence of the numerically computed eigenfrequencies with respect to the truncation order $j_\mathrm{max}$ in the Taylor expansions~\eqref{eq:Taylor_expand} is assessed through the quantities $\lvert E_{011}/E_{011}^{j_\mathrm{max}=0} - 1\rvert$ and $\lvert \Gamma_{011}/\Gamma_{011}^{j_\mathrm{max}=0} - 1\rvert$, as shown in Fig.~\ref{fig:vs_jmax}. The figure demonstrates stable convergence for $j_{\rm max}\gtrsim 6$. Therefore, we conservatively adopt $j_\mathrm{max}=10$ for all computations in the main text regarding the disk model.

For completeness, in Figs.~\ref{fig:diff_epsilon} and~\ref{fig:diff_mu_disk} we also show how the perturbed quasibound-state spectrum varies with $\epsilon_{\rm d}$ and $\alpha$, respectively. Overall, the general behavior confirms the one obtained in the galactic halo case in the main text, with the relativistic covariant formulation providing the best approximation overall. 

\section{Expansion coefficients within the completeness assumption}
\label{appendix:linear_system_derivation}

In this appendix we provide a detailed derivation of the linear system that determines the expansion coefficients $c_{\mathbf{p}}$ in Eq.~\eqref{eq:chi_spatial_expansion}.

Substituting the perturbative expansions of $\omega_{nlm}$, $\Phi_{nlm}$ and $\mathcal{O}$ into the Klein-Gordon equation and collecting terms order by order, the first-order equation reads
%%%%%=====
\begin{align}\label{eq:KG_first_order_app}
    \mathcal{O}^{(0)}\,\delta\Phi_{nlm}^{(1)} + \delta\mathcal{O}^{(1)}\Phi_{nlm}^{(0)} = 0.
\end{align}
%%%%%
Since the mode $\Phi_{nlm}$ satisfies $\partial_t\Phi_{nlm}=-i\omega_{nlm}\Phi_{nlm}$, expanding this relation at first order yields
%%%%%=====
\begin{align}\label{eq:Lie_first_order_app}
    \partial_t\,\delta\Phi_{nlm}^{(1)} = -i\omega_{nlm}^{(0)}\,\delta\Phi_{nlm}^{(1)} - i\delta\omega_{nlm}^{(1)}\,\Phi_{nlm}^{(0)},
\end{align}
%%%%%
whose general solution is precisely the ansatz given in Eq.~\eqref{eq:first_order_mode_time_dependence}.

Inserting Eq.~\eqref{eq:first_order_mode_time_dependence} into Eq.~\eqref{eq:KG_first_order_app} the time-domain operator $\mathcal{O}^{(0)}$ produces two pieces: on the non-secularly growing component, the time-domain operator simply produces its frequency-domain version $\mathcal{O}^{(0)}_{\rm freq}(\omega_{nlm}^{(0)})$ via $\partial_t^2\rightarrow-(\omega_{nlm}^{(0)})^2$; on the secularly-growing component, the second derivative $\partial_t^2$ contained in $\mathcal{O}^{(0)}$ acts on the explicit $t$ factor and produces an additional finite contribution. Using the zeroth-order equation $\mathcal{O}^{(0)}_{\rm freq}(\omega_{nlm}^{(0)})\chi_{nlm}^{(0)} = 0$, the $t$-linear pieces cancel exactly. The result is the time-independent spatial equation for $\chi_{nlm}^{(1)}$,
%%%%%=====
\begin{align}\label{eq:chi_eq_app}
    \begin{split}
        \mathcal{O}^{(0)}_{\rm freq}(\omega_{nlm}^{(0)})\chi_{nlm}^{(1)} + \delta\mathcal{O}_{\rm freq}^{(1)} &(\omega_{nlm}^{(0)}) \chi_{nlm}^{(0)} 
        \\ 
        &- \frac{2\omega_{nlm}^{(0)}\delta\omega_{nlm}^{(1)}}{f(r)}\chi_{nlm}^{(0)} = 0,
    \end{split}
\end{align}
%%%%%
where the last term originates from the action of $\partial_t^2$ on the secular component of $\delta\Phi_{nlm}^{(1)}$. Equivalently, it corresponds to the $\delta\omega_{nlm}^{(1)}\partial\mathcal{O}^{(0)}_{\rm freq}/\partial\omega$ contribution arising when the eigenfrequency is expanded inside the unperturbed operator. Although the secular component grows linearly with time, this cancellation ensures that the spatial equation for $\chi_{nlm}^{(1)}$ is itself time-independent.

We now multiply Eq.~\eqref{eq:chi_eq_app} by $f(r)$, which puts the operator in the form whose natural weight matches the Schwarzschild bilinear form. Recalling $\mathcal{F}_{nlm}\equiv f(r)\delta\mathcal{O}_{\rm freq}^{(1)}(\omega_{nlm}^{(0)})$ from the main text, this gives
%%%%%=====
\begin{align}\label{eq:f_chi_eq_app}
    \begin{split}
        f(r)\mathcal{O}^{(0)}_{\rm freq}(\omega_{nlm}^{(0)})\chi_{nlm}^{(1)} + \mathcal{F}_{nlm}&\,\chi_{nlm}^{(0)} \\
        &- 2\omega_{nlm}^{(0)}\delta\omega_{nlm}^{(1)}\,\chi_{nlm}^{(0)} = 0.
    \end{split}
\end{align}
%%%%%
Since the zeroth-order equation $\mathcal{O}^{(0)}_{\rm freq}(\omega_{\mathbf{m}}^{(0)})\chi_{\mathbf{m}}^{(0)}=0$ holds for every basis mode and $\mathcal{O}^{(0)}_{\rm freq}(\omega)$ depends on $\omega$ only through the multiplicative term $-\omega^2/f(r)$, we obtain the useful identity
%%%%%=====
\begin{align}\label{eq:identity_app}
    f(r)\mathcal{O}^{(0)}_{\rm freq}(\omega_{nlm}^{(0)})\chi_{\mathbf{m}}^{(0)} = \left[(\omega_{\mathbf{m}}^{(0)})^2 - (\omega_{nlm}^{(0)})^2\right]\chi_{\mathbf{m}}^{(0)}.
\end{align}
%%%%%
Substituting the basis expansion~\eqref{eq:chi_spatial_expansion} into Eq.~\eqref{eq:f_chi_eq_app} and using Eq.~\eqref{eq:identity_app} converts the differential equation into the algebraic relation
%%%%%=====
\begin{align}\label{eq:first_order_with_extra_app}
    \begin{split}
        \sum_{\mathbf{m}\in\mathcal{B}_{nlm}} c_{\mathbf{m}} & \left[(\omega_{\mathbf{m}}^{(0)})^2 - (\omega_{nlm}^{(0)})^2\right]\chi_{\mathbf{m}}^{(0)} 
        \\
        &+ \mathcal{F}_{nlm}\chi_{nlm}^{(0)} - 2\omega_{nlm}^{(0)}\delta\omega_{nlm}^{(1)}\chi_{nlm}^{(0)} = 0.
    \end{split}
\end{align}
%%%%%
Applying the projector $\mathcal{P}_{\mathbf{p}}$ defined in the main text [see Eq.~\eqref{eq:projetor}] to Eq.~\eqref{eq:first_order_with_extra_app} for each $\mathbf{p}\in\mathcal{B}_{nlm}$ and using the orthogonality property in~\eqref{eq:ortho_main}, the sum over $\mathbf{m}$ collapses to its single $\mathbf{m}=\mathbf{p}$ term. Moreover, since $\mathbf{p}\notin\{nlm\}$, the projection $\mathcal{P}_{\mathbf{p}}[\chi_{nlm}^{(0)}]=\mathcal{N}_{nlm}\delta_{\mathbf{p},nlm}=0$, so the last term in Eq.~\eqref{eq:first_order_with_extra_app} (originating from the secular component of $\delta\Phi_{nlm}^{(1)}$) drops out of every projection. The coefficient equation thus immediately reduces to Eq.~\eqref{eq:c_decoupled_main} of the main text.

\bibliography{refs}% Produces the bibliography via BibTeX.  

\end{document}